\documentclass[11pt]{article}
\pdfoutput=1

\usepackage[english]{babel}
\usepackage{fancyhdr}
\usepackage{amsmath}
\usepackage{amssymb}
\usepackage{amsfonts}
\usepackage{psfrag}
\usepackage[applemac]{inputenc}
\usepackage{graphicx,wrapfig}
\usepackage{pstricks}

\usepackage{slashed}            
\usepackage{subfig}             
\usepackage{xspace}				
\usepackage[sort&compress,square,comma,numbers]{natbib}
\usepackage{float}
\usepackage{dcolumn}
\usepackage{bm}
\usepackage{epstopdf}

\usepackage[colorlinks=true,
citecolor=blue,
linkcolor=black,
urlcolor=blue,
hypertexnames=false]{hyperref}

\numberwithin{equation}{section}
	
\DeclareFontFamily{OT1}{pzc}{}
\DeclareFontShape{OT1}{pzc}{m}{it}{<-> s * [1.100] pzcmi7t}{}
\DeclareMathAlphabet{\mathpzc}{OT1}{pzc}{m}{it}

\newcommand{\Tr}{\mathop{\rm Tr}}
\def\p{{\bf{p}}}
\def\k{{\bf{k}}}

\newcommand{\beq}{\begin{eqnarray}}
\newcommand{\eeq}{\end{eqnarray}}
\newcommand{\bea}{\begin{eqnarray}}
\newcommand{\eea}{\end{eqnarray}}
\newcommand{\bag}{\begin{align}}
\newcommand{\eag}{\end{align}}

\begin{document}

\baselineskip=18pt

\setcounter{footnote}{0}
\setcounter{figure}{0}
\setcounter{table}{0}


\begin{titlepage}

\hfill{Saclay-t16/038}
\vspace{1cm}

\begin{center}
  \begin{Huge}
     \begin{bf} Softness and Amplitudes' Positivity \\ \vspace{0.3cm} for Spinning Particles \vspace{0.3cm}   \end{bf}
  \end{Huge}
\end{center}
\vspace{0.4cm}
\begin{center}
\begin{large}
{\bf Brando Bellazzini}
\end{large}
  \vspace{0.4cm}
  
  \begin{it}
\begin{small}
Institut de Physique Th\'eorique, Universit\'e Paris Saclay, CEA, CNRS, F-91191 Gif-sur-Yvette, France
\vspace{0.3cm}\\

Dipartimento di Fisica e Astronomia, Universit\'a di Padova, 
Via Marzolo 8, I-35131 Padova, Italy

\end{small}

\end{it}

\end{center}

\begin{abstract}
\medskip
\noindent

We derive positivity bounds for scattering amplitudes of particles with arbitrary spin using unitarity, analyticity and crossing symmetry.  
The bounds imply the positivity of certain low-energy coefficients of the effective action that controls the dynamics of the light degrees of freedom. We show that 
  low-energy amplitudes strictly softer than $O(p^4)$ do not admit unitary ultraviolet completions unless the theory is free. 
This enforces  a bound on the energy growth of scattering amplitudes in the region of validity of the effective theory. 
 We discuss explicit examples including the Goldstino from spontaneous supersymmetry breaking, and the theory of a spin-1/2 fermion with a shift symmetry.

\end{abstract}

\bigskip

\end{titlepage}

\tableofcontents

\setcounter{equation}{0}
\setcounter{footnote}{0}

\section{Introduction} \label{sec:intro}

Effective Field Theories (EFT's) describe the dynamics of light degrees of freedom at low-energy via higher dimensional operators that incapsulate the effect of high-energy physics which is kinematically inaccessible. 
Symmetries play a central role in the study of EFT's as the renormalization group flow from the ultraviolet (UV) to the infrared (IR) fixed points respects them.
 As a matter of fact the converse is essentially true as well, and operators that are not protected by a symmetry are expected to be generated along the renormalisation group flow: ``write down all terms allowed by the symmetries''  is the mantra of EFT practitioners. Symmetries, either approximate or exact, provide also an organising principle, or  power counting, that determines which operator is important and which one is instead naturally suppressed by insertions of small spurions. 

But besides symmetries there are other, perhaps more structural, conditions which an EFT must rely upon, and that further constrain the structure of the low-energy theory.  
It is well known that not every EFT in the IR admits an UV completion consistent with the fundamental $S$-matrix properties \cite{Adams:2006sv}.  Crossing symmetry and the analytic structure of scattering amplitudes of the underlying microscopic theory provide a link between the UV- and the IR-theory in the form of dispersion relations for the elastic forward $2\rightarrow 2$ scattering $\phi X\rightarrow \phi X$ \cite{Weinberg:1995mt}. 
These relations provide a UV-IR connection because the low-energy amplitude in the deep IR is expressed as a dispersive integral of the discontinuity across the branch cuts which extend to arbitrary high-energy in the complex $s$-plane. 
Moreover, unitarity of the microscopic theory implies the optical theorem that insures strictly positive discontinuities across the branch cuts at all energies, and in turn the positivity constraints of the schematic form 
\begin{equation}
\label{Eq:intropos}
\frac{\partial^{2}}{\partial s^2}\mathcal{M}^{EFT}(\phi X\rightarrow \phi X)\big|_{s=t=0}>0
\end{equation}
on the low-energy amplitudes \cite{Adams:2006sv}. 

The  positivity constraints do not depend on the specific dynamics of the UV completion, as they are obtained by fundamental requirements such as unitarity, crossing symmetry and analyticity that are usually assumed in any scattering theory. This is why they have found several applications ranging e.g. 
from the a-theorem \cite{Komargodski:2011vj} to the theory of pions \cite{Pham:1985cr,Manohar:2008tc,Mateu:2008gv}, from WW-scattering \cite{Adams:2006sv,Distler:2006if,Bellazzini:2014waa} to composite Higgs models \cite{Bellazzini:2014waa,Low:2009di,Falkowski:2012vh,Urbano:2013aoa}, from quantum gravity \cite{Bellazzini:2015cra} to inflation \cite{Baumann:2015nta,Croon:2015fza}, from Galileons \cite{Nicolis:2009qm} to massive gravity \cite{Cheung:2016yqr}, from the weak gravity conjecture \cite{Cheung:2014ega,Cheung:2014vva} to the OPE coefficients \cite{Komargodski:2016gci}, from the conformal blocks expansion \cite{Hartman:2015lfa} to the Mellin amplitudes for CFTs at large-N \cite{Alday:2016htq}. 
Virtually all literature have focused on positivity bounds for amplitudes of bosons with spin-0, -1 or -2; see \cite{Adams:2008hp}  for an interesting exception that studied dimension-6 4-fermi interactions, and  e.g.  \cite{Goldberger:1955zza,Hamilton:1963zz,Luo:2006yg,Sanz-Cillero:2013ipa}  for pion-nucleon scattering in QCD.
And in fact, apart from e.g. \cite{Bellazzini:2015cra,Cheung:2016yqr}, most of the positivity bounds for massive spin-1 or spin-2 bosons have actually focused on the ``eaten'' scalar modes, i.e. the spin-0 Goldstone Bosons (GB) or the Galileon mode, respectively. 

In this paper we close this gap and study in generality the positivity bounds  for scattering amplitudes $\mathcal{M}(\chi^{\sigma_1} \psi^{\sigma_2}\rightarrow \chi^{\sigma_1} \psi^{\sigma_2})$ between particles $\chi$ and $\psi$ of \textit{arbitrary} spins (or helicities) $\sigma_i$, including the case where both are fermions with half-integer spins.

This task is non-trivial as the polarizations for spinning particles depend on the momenta and carry themselves discontinuities in the complex $s$-plane that could affect the analytic structure of the whole amplitude which is built out of the amputated correlators dotted with the polarizations. 
Moreover, crossing symmetry for spinning particles is not, in general, as simple as exchanging the Mandelstam's variables $s\leftrightarrow u$. Furthermore, crossing fermions introduces an extra minus sign relative to the otherwise identical prescription for bosons, and one must ensure that this sign does not propagate and spoil the positivity of the integrand in the dispersion relation.
 In fact, we show that  crossing symmetry and the polarizations  for spinning particles actually conspire together, but only in the special kinematics of the forward limit, to cancel each other's issues  and yield again the positivity bounds (\ref{Eq:intropos}).  The resulting positivity  bounds can thus be elevated to universal statements about scattering amplitudes\footnote{Certain caveats apply; they are discussed in section~\ref{sec:unitarity}.}.

An interesting consequence of these positivity bounds is that scattering amplitudes can not be arbitrarily soft. They can be as soft as $O(p^4)$, but not softer\footnote{In this paper by \textit{soft} it is  meant that \textit{all} external momenta in the scattering are sent to zero with the same scaling factor.}. In turn this imposes a constraint on how quickly amplitudes can raise with energy within the validity of the EFT.
Particles that are highly boosted but with momenta below the cutoff of the EFT, $m\ll E \ll \Lambda$, can be considered massless and soft compared to $\Lambda$.  The leading soft behavior tells us that amplitudes with energy above the IR thresholds, but within the validity of the EFT, can be dominated at most by an $O(p^4)$-behavior.  The EFT contains of course higher energy corrections that come from higher-dimensional operators but they are always subleading because suppressed by positive powers of $E/\Lambda\ll 1$, relative to the leading soft behavior. Those corrections would  become important only at the cutoff $E\sim \Lambda$ where the EFT breaks down.  
All in all, fundamental properties of scattering amplitudes enforce the $O(p^4)$-limit on the leading energy growth behavior of the amplitudes for particle with arbitrary spin, and within the validity of the EFT. 

In section~\ref{sec:softlimi} we discuss the example of a spin-1/2 chiral fermion that saturates this soft behavior, reproducing essentially  the structure of the theory of the Goldstino from spontaneous breaking of $N=1$ supersymmetry (SUSY). 

Even though the amplitudes can not be strictly softer than $O(p^4)$ one may wonder whether there exist a loose sense or an approximate limit in which they can effectively be softer than that, i.e.  \textit{supersoft}.
It is in principle conceivable a theory of massless particles where the amplitudes are as soft as $O(p^6)$ and yet respect all our consistency requirements whenever a tiny, in fact arbitrarily small, mass or any another IR scale $\Lambda_{IR}$, is  generated by a less irrelevant perturbation than those responsible for the leading interactions suppressed by the scale $\Lambda$ that controls the derivatives expansion. The perturbation gives rise to a tiny yet important correction of $O(\Lambda_{IR}^2 p^4/\Lambda^6)$ which allows the amplitudes to satisfy the positivity constraints for arbitrarily small values of $\Lambda_{IR}$.  This happens e.g. for the theory of massive gravity where the leading Galileon mode scales as $O(p^6)$ while the sub-leading contributions go like $m_g^2 p^4/\Lambda^6$ \cite{Cheung:2016yqr} and $\Lambda_{IR}$ is identified with the graviton mass $m_g$.  

We discuss similar theories in section~\ref{sec:nonunitary} and make the non-decoupling between UV and IR more manifest. As we remove the IR deformation, i.e. $\Lambda_{IR}\rightarrow 0$, we hold  fixed  the coupling $g_*$ that is controlling the supersoft terms at a certain energy scale $\Lambda$ that eventually diverges,  
$\Lambda\rightarrow \infty$, but arbitrarily slowly.
 In this way the supersoft $O(p^6)$-terms fail to dominate the amplitude at low-energy only in a tiny window in the deep IR, $0<E < \Lambda_{IR}$, that can be shrunk to zero much faster than the rate by which the amplitudes, say $\mathcal{M}(\chi\psi\rightarrow \chi\psi)\sim g_*^2 (E/\Lambda)^6$, eventually vanish as $\Lambda\rightarrow \infty$.
We can make the hierarchy of scales $\Lambda_{IR}\ll \Lambda$ completely natural by a symmetry which is broken by a small spurion $\epsilon$ so that $\Lambda_{IR}\sim \epsilon \Lambda$.   Equivalently,  we can consider hierarchical couplings  $g_*$ and $g= \epsilon g_*$, which control the supersoft and the $O(p^{4})$-behavior of the amplitudes  respectively,  because of an enhanced   symmetry for $\epsilon\rightarrow 0$. 
Consistency with our positivity bounds is obtained when $\epsilon\rightarrow 0$ (and therefore $\Lambda_{IR}\,,g \rightarrow 0$) by demanding that $\Lambda$ weakly depends on $\epsilon$ such that  nevertheless  $\Lambda\rightarrow \infty$, eventually. This dependence is taken weak arbitrarily, say e.g.  $\Lambda\sim \log\epsilon $ or $\Lambda\sim \epsilon^{-n}$ with $n\ll1$.  For any finite but small $\epsilon$ we get essentially a supersoft theory to almost all energies below the cutoff.
We argue in section~\ref{obstLoop} that, however, other obstructions may forbid taking $\epsilon$ very small while the coupling $g_*$ of the UV completion is held finite. 

We study in detail an example of fermionic supersoft theory: a spin-1/2 particle with a fermionic shift symmetry $\chi\rightarrow \chi+\xi$ which is perturbed by Goldstino-like (less) irrelevant interactions that can be suppressed by the spurion $\epsilon$ associated with the breaking of the shift symmetry. The IR scale $\Lambda_{IR}$ can be essentially identified with $\epsilon \Lambda$ where $\Lambda$ is the cutoff that suppresses the interactions respecting the shift symmetry.
The hierarchy between $\Lambda$ and $\Lambda_{IR}$ is naturally stable since it is controlled by approximate symmetries. 
This example provides an even steeper energy growth than the one proposed with the ``remedios''  in \cite{Liu:2016idz}. 
The remedios, as well as the supersoft theories, are interesting because new-physics effects that would become visible at high-energy disappear quickly going to lower energy. This makes them consistent with the absence of new physics in the IR.  Going to higher energy, certain higher dimensional operators quickly overrun the amplitudes and dominate the lower dimensional operators that are  suppressed by the small spurions, while remaining within the regime of validity of the EFT. In this sense, these theories run very fast while the amplitude remains small. The relevance of unsuppressed higher dimensional operators that produce such stronger dependence upon the energy scale, and may hence affect the phenomenology at the Large Hadron Collider, has been recently emphasised e.g. in \cite{Contino:2016jqw}.

The paper is organised as follows: we first review crossing symmetry in section~\ref{sec:crossing}, we study the analytic properties of the polarizations and their relation to crossing symmetry in section~\ref{sec:frowardAndCross}, we derive the general positivity bounds in section~\ref{sec:unitarity}, we study the maximally soft theory of a spin-1/2 fermion in section~\ref{sec:softlimi}, we discuss  supersoft theories in section~\ref{sec:nonunitary}, and we finally conclude in section~\ref{sec:conc}.

\section{Crossing Symmetry} \label{sec:crossing}

In this section we recall the basic properties of crossing symmetries and introduce the notation used throughout the paper. 
The impatient reader may go directly to section~\ref{sec:frowardAndCross}. 

\subsection{Crossing one particle}

\begin{figure}[thb]
\centering
\includegraphics[width=8cm]{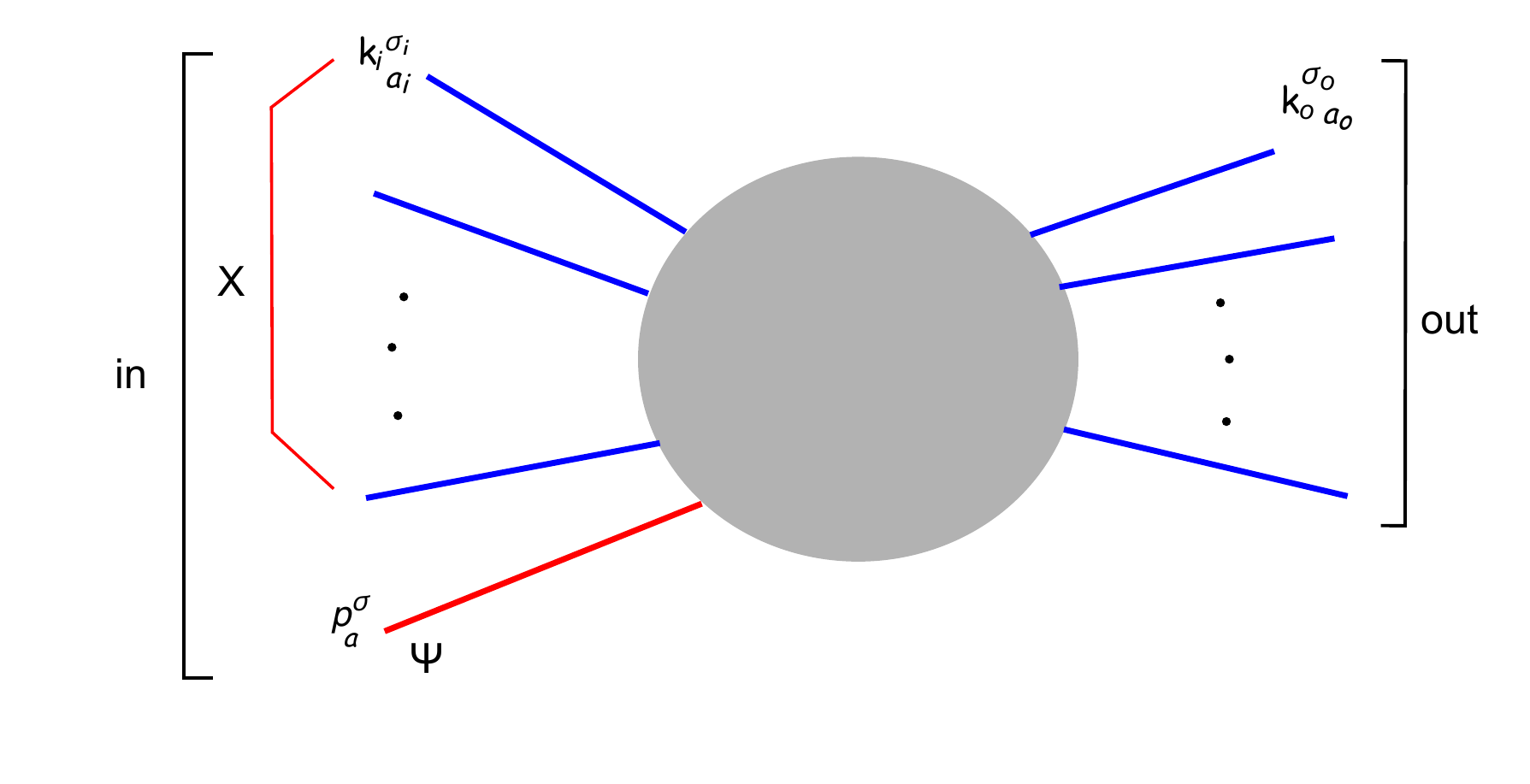}
\includegraphics[width=8cm]{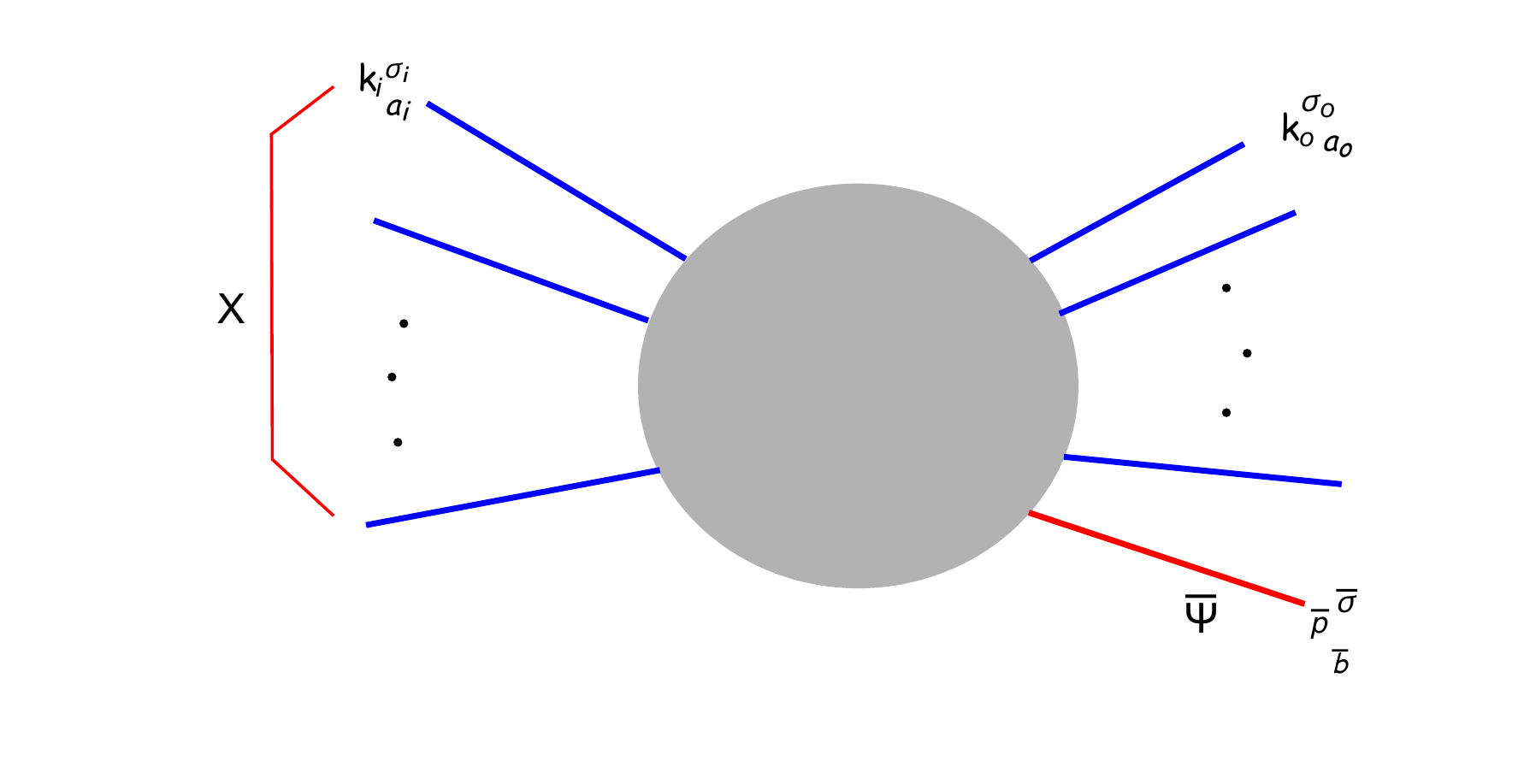}
\caption{Scattering processes $X+\Psi \rightarrow \mathrm{out}$ and $X\rightarrow \left(\mathrm{out}+\overline\Psi\right)$ which are related by crossing symmetry according to Eq.~(\ref{scattering0}) and (\ref{scattering0cross}).}
\label{fig:1}
\end{figure}

Consider a  scattering process $in\rightarrow out$ represented by the transition
$$
\{k^{\sigma_i}_{i\, a_i},p^\sigma_a\} \longrightarrow \{k^{\sigma_o}_{o\, a_o}\}
$$
where a certain particle $\Psi$  of 4-momentum $p$, little-group index $\sigma$ (either the spin or the helicity), and internal index $a$ belongs to the initial initial state $in=\Psi+X$ together with other particles, that may or may not be of the same species of $\Psi$, which are collectively called $X$, see Fig.~\ref{fig:1}. 
The $in$ and $out$ quantum numbers are collectively represented by the set of quantum numbers $\{k^{\sigma_{i}}_{i\, a_{i}}, p^\sigma_a\}$ and $\{k^{\sigma_{o}}_{o\, a_{o}}\}$ respectively.

The internal indexes $a_j$ label the species of the particles and, possibly, the elements inside the representation $\mathbf{r}_j$ of the symmetry group carried by the states. 
For massless particles the index $\sigma$ can be identified with the helicity, whereas for massive states it takes the $2S+1$ values from $-S$ to $S$, where $S$ is the spin. 
  It may be sometimes convenient to work with states of non-definite helicity or spin, such as e.g. photons with linearised rather than circular polarizations. For the rest of this section we focus only on states of definite spin or helicity, leaving the discussion of linear polarizations to appendix~\ref{sec:crossforw}. 
  
The scattering amplitude takes the form 
\begin{equation}
 \label{scattering0}
 \mathcal{M}(\{k_{i\,a_i}^{\sigma_i}, p^{\sigma}_a \}  \rightarrow \{k_{o\,a_o}^{\sigma_o}\}) =\mathcal{O}^{\ell}(\{k^{\sigma_i}_{i\, a_i},k^{\sigma_o}_{o\,a_o}\}, p;a) u^{\sigma}_{\ell}(\p)
\end{equation}
where we have singled out the polarization $u^{\sigma}_{\ell}(\p)$ of $\Psi$. 
The polarizations $u^\sigma_\ell$ ($v^\sigma_{\ell}$) for (anti-)particles carry a little group index $\sigma$ as well as a Lorentz or spinorial index $\ell$.~\footnote{For a generic representation of the Lorentz group, the index $\ell$ is actually a pair of indexes $(a,b)$ that labels the $(2A+1)(2B+1)$ states in the irreducible representations $(A,B)$ of $SU(2)\times SU(2)\sim SO(3,1)$ identified by half-integers values for $A$ and $B$. The $(0,0)$, $(1/2,0)$, $(0,1/2)$, $(1/2,0)\oplus (0,1/2)$, and $(1/2,1/2)$ are the standard scalar, left-handed spinor, right-handed spinor, Dirac spinor, and vector representation respectively. The Dirac and vector representations are often recast with a single index, $\ell=\alpha$ for the 4-component Dirac spinors $u^\sigma_\alpha$ and $v^\sigma_\alpha$, and $\ell=\mu$ for the 4-vector polarizations $\epsilon^\sigma_\mu$ and $\epsilon^{\sigma\,*}_\mu$.} 
  Contracting the polarizations  with the Lorentz-covariant  residues of the $n$-point correlation functions, i.e. with the on-shell matrix elements amputated by the external propagators  $ \Delta_{\ell \ell^\prime}(k)$, schematically e.g.
\begin{equation}
\label{corrrelation}
\Delta^{-1}_{\ell\ell^\prime}(p) \prod_{i,o} \Delta^{-1}_{\ell_i \ell_i^\prime}(k_i)  \Delta^{-1}_{\ell_o \ell_o^\prime}(k_o) \int d^4 x_1\ldots d^4 x e^{-ipx -i\sum k_i x_i+k_o x_o} \langle 0|\mathrm{T}\, X_{\{\ell^\prime_i\ell^\prime_o\}}(\{x_i, x_o\}) \Psi^{\dagger}_{\ell^\prime\, a}(x) |0\rangle \,,
\end{equation}
 we extract the little-group covariant on-shell S-matrix elements \cite{Weinberg:1995mt}.  This is nothing but the standard LSZ reduction formula \cite{Weinberg:1995mt,Srednicki:2007qs} where $X_{\{\ell_i\ell_o\}}(\{x_i, x_0\}) $ is a shorthand for the product of fields $\Phi_{a_i\,\ell_i}(x_i)$ which are Fourier transformed with momenta $+k_i$ if ingoing and $-k_o$ if outgoing, and that have a non-vanishing overlap between the vacuum and the one-particle states occurring in the scattering. We call  $\Psi_{\ell\,a}(x)$ any field that annihilates  the particle $\Psi$ 
\begin{equation}
\label{overlap}
\langle 0|\Psi_{\ell\,b}(x) |p^\sigma_a \rangle \propto \delta_{ab}  u^\sigma_{\ell}(\p) e^{-ipx}
\end{equation}
and that creates  its  anti-particles $\overline{\Psi}$
\begin{equation}
\label{overlapLoc}
\langle p^{\sigma}_{\bar{a}}|\Psi_{\ell\, b}(x) |0\rangle \propto \delta_{ab}  v^{\sigma}_{\ell}(\p)e^{ipx}\,.
\end{equation}
The proportionality factor is an irrelevant constant that is removed by the LSZ reduction formula\footnote{We are working with the relativistic normalization $\langle p^{\sigma} | k^{\sigma^\prime}\rangle =\delta^{\sigma\sigma^\prime}(2\pi)^3 2E_\p \delta^3(\p-\k)$.}. 
Anti-particles are denoted by a bar over the internal index which labels the the states inside the complex conjugate representation $\mathbf{r}^*$ carried by $\overline\Psi$. For a $U(1)$ $a$ is the charge and $\bar{a}=-a$.

The scattering amplitude (\ref{scattering0}) is related via crossing symmetry to another physical process where the antiparticle $\overline\Psi$ of generic momentum $\overline{p}$, helicity (or spin) $\overline{\sigma}$, and internal quantum number $\overline{b}$ belongs to the final state, while the other particles in the $in$ and $out$ state have not been touched, i.e. $X\rightarrow \left(\mathrm{out}+\overline\Psi\right)$, see Fig.~\ref{fig:1}.
This  \textit{crossed amplitude} reads
\begin{equation}
 \label{scattering0cross}
 \mathcal{M}(\{k_{i\,a_i}^{\sigma_i}\}  \rightarrow  \{k_{o\,a_o}^{\sigma_o}, \overline{p}^{\overline{\sigma}}_{\overline{b}} \} ) = \pm \,\mathcal{O}^{\ell}(\{k^{\sigma_i}_{i\, a_i},k^{\sigma_o}_{o\,a_o}\}, -\overline{p}; b) \,v^{\overline{\sigma}}_{\ell}(\overline{\p})
\end{equation}
where  $\mathcal{O}^{\ell}(\{k^{\sigma_i}_{i\, a_i},k^{\sigma_o}_{o\,a_o}\}, -\overline{p}; b) $ is the same function that appears in Eq~(\ref{scattering0}) but evaluated at the unphysical momentum $-\overline{p}$ given that $\Psi$ belongs to the final states. Equivalently, the pole at $p^2=m_\Psi^2$ originates from the overlap (\ref{overlapLoc}) rather than (\ref{overlap}).
The overall sign is determined by the statistics of $\Psi$: it's $+$ for a boson and $(-1)^n$ for a fermion, where $n$ is the number of fermion pairs exchanges we need to perform to move $\Psi_\ell(x)$ through the fields $\Phi_{a_i\,\ell_i}$ in $X_{\ell_1\ell_2,\ldots}(\{x_i, x_o\})$ in order to reach the canonical form $\langle 0|\mathrm{T}\, \Psi^{\dagger}_{\ell\, a}(x) X_{\ell_1\ell_2,\ldots}(\{x_i, x_o\}) |0\rangle$ in the amputated matrix element. 

The polarization dotted with the amputated matrix element is now $v^{\overline{\sigma}}_\ell$ rather than $u^\sigma_\ell$ because the relevant overlap is given by Eq.~$(\ref{overlapLoc})$ as opposed to Eq.~$(\ref{overlap})$. Crucially enough, the particle/anti-particle polarizations are actually related  by locality and causality that enforce, via $CPT$ invariance, the following relations
\begin{align}
\label{eq:uvgamma5}
\epsilon^{\sigma\,*}_\mu(\p)=(-1)^{\sigma}\epsilon^{-\sigma}_\mu(\p)\,,  \qquad v^{\pm}(\p)=\mp \gamma^5 u^{\mp}(\p)\,,\qquad v^{+}_{L}(\p)=u^-_{L}(\p)\,,\qquad v^{-}_{R}(\p)=u^{+}_{R}(\p)\,,
\end{align}
 for vector, Dirac, and left- or right-handed (massless) Weyl representation respectively\footnote{In order to avoid clutter of notation we display  only the sign $\pm$ for the spin-1/2 label $\sigma$, which is shorthand for the actual value equal to $\pm 1/2$. Moreover, the overall sign that relates $u^\sigma$ with $v^{-\sigma}$ in (\ref{eq:uvgamma5}) is conventional since it depends on the choice of the $CPT$ phase. In contrast, the resulting relation (\ref{Eq:massiverhosrels}) for the density matrices $\rho^\sigma=u^\sigma(\p) u^{\sigma\,\dagger}(\p)$ and $\widetilde{\rho}^{-\sigma}=v^{-\sigma}(\p) v^{-\sigma\,\dagger}(\p)$ is physical and does not depend on any conventional choice.}.  Therefore, simple relations emerge between the scattering amplitude and  its crossed amplitude where the helicities of the crossed particles are reversed, $\bar{\sigma}=-\sigma$. Actually, massless Weyl particle/anti-particles have opposite helicity anyway. 
The proof of Eq.~(\ref{eq:uvgamma5}), as well as its generalisation~(\ref{eq:generalCPTwave}) to arbitrary representations $(A,B)$ of the Lorentz group $SO(3,1)\sim SU(2)\times SU(2)$, is discussed in appendix~\ref{app:polarizations}. 

\subsection{Crossing two or more particles}

It is clear now how to extend the action of crossing to more particles that swing side between the $in$ and the $out$ state. 
 For any particle with quantum numbers $p^\sigma_a$ in the initial state which is replaced by its anti-particle of quantum numbers $\overline{p}^{\overline{\sigma}}_{\overline{a}}$ in the final state\footnote{Since the anti-particles are denoted by $\overline{\Psi}$ relative to the particles $\Psi$, their generic 4- and 3-momentum are also called $\overline{p}$ and $\bar{\p}$,  analogously to the notation for the charges and the spins. } we keep the same amputated matrix element of the original process but evaluated at $p\rightarrow -\overline{p}$, we multiply it by an overall sign determined by the statistic, and we replace the polarizations $u^\sigma(\p)\leftrightarrow v^{\bar{\sigma}}(\bar{\p})$.  Of course, for (anti-)particles in the final (initial) state that move to the other side of the scattering, the replacement is $u^{\sigma\dagger}(\p)\leftrightarrow v^{\bar{\sigma}\dagger}(\bar{\p})$. Notice that the polarizations are always on-shell, i.e. $p^0=\sqrt{\p^2+m_\Psi^2}$ and $\bar{p}^0=\sqrt{\bar{\p}^2+m_\Psi^2}$, i.e. their 4-momentum is not flipped under crossing.

For example, let us consider the scattering 
$$\Psi(k_{1\,a_1}^{\sigma_1}) X(k_{i\, a_i}^{\sigma_i}) \longrightarrow \Psi(k_{3\,a_3}^{\sigma_3}) X(k_{o\,a_o}^{\sigma_o})$$
 where particles $1$ and $3$ are of the same species.   This scattering includes the simple $2\rightarrow 2$ scattering as a special case where $i=2$ and $o=4$. 
We consider the crossed process 
$$\overline{\Psi}(\overline{k}_{1\,\overline{a}_1}^{\overline{\sigma}_1}) X(k_{i\, a_i}^{\sigma_i}) \longrightarrow \overline{\Psi}(\overline{k}_{3\,\overline{a}_3}^{\overline{\sigma}_3}) X(k_{o\,a_o}^{\sigma_o})$$ 
where we have crossed particles $1$ and $3$.
The scattering amplitudes are given by\footnote{It is customary for Dirac spinors to work with $\bar{u}^\sigma \equiv u^{\sigma\,\dagger}\gamma^0$ and $\bar{v}^{\sigma} \equiv v^{\sigma\,\dagger} \gamma^0$ by absorbing an extra $\gamma^0$ in the amputated matrix elements. For our purposes we find instead more convenient in the following to work always with $u^\dagger$ and $v^\dagger$.}
\begin{align}
\label{eq:crossed2bodyone}
  \mathcal{M}(  k_{1\,a_1}^{\sigma_1}, k_{2\,a_2}^{\sigma_2}  \rightarrow   k_{3\,a_3}^{\sigma_3},k_{4\,a_4}^{\sigma_4}     ) =  & \left[\ldots u^{\sigma_3\,*}_{\ell_3}(\k_3)\right] \mathcal{O}_{\ell_3\ldots}^{\ell_1\ldots }(k_1, k_3, \ldots ;a_1, a_3\ldots) \left[u^{\sigma_1}_{\ell_1}(\k_1) \ldots \right]  \\
\label{eq:crossed2body}
 \mathcal{M}(  \overline{k}_{1\,\overline{a}_1}^{\overline{\sigma}_1}, k_{2\,a_2}^{\sigma_2}    \rightarrow   \overline{k}_{3\,\overline{a}_3}^{\overline{\sigma}_3}, k_{4\,a_4}^{\sigma_4},    ) = &  \eta_{\Psi} \left[\ldots v^{\bar{\sigma}_1\,*}_{\ell_3}(\bar{\k}_1)\right]  \mathcal{O}_{\ell_3\ldots}^{\ell_1 \ldots}(-\overline{k}_3, -\overline{k}_1, \ldots; a_3, a_1\ldots )\left[v^{\bar{\sigma}_3}_{\ell_1}(\bar{\k}_3) \ldots\right]\,.
\end{align}
where the $\ldots$ hides the irrelevant dependence on the spectators' polarizations and quantum numbers, that is the dependence on particles $2$ and $4$.
The overall sign $\eta_\Psi=(-1)^{2S_\Psi}$ is $+(-)$ for $\Psi=$~boson(fermion), and it does not depend on the spectators $X$ as the amplitude must involve an even number of fermions and hence an odd number of fermion exchanges while keeping the same ordering for the $X's$.

\section{Forward scattering and crossing as $s\leftrightarrow u$}
\label{sec:frowardAndCross}

The $2\rightarrow 2$ scattering amplitude of spin-0 particles $$\pi^{a_1}(k_1) \pi^{a_2}(k_2)\longrightarrow \pi^{a_3}(k_3) \pi^{a_4}(k_4)$$
 such as pions of QCD is a little-group singlet that transforms as a scalar under a Lorentz transformation $\Lambda$, implying $\mathcal{M}(\{p_i\})=\mathcal{M}(\{\Lambda p_i\})$.  
 The amplitude is Lorentz invariant and can be written in terms of Mandelstam's variables alone, i.e. $\mathcal{M}_{a_1 a_2 a_3 a_4}(s,t,u)$, where 
\begin{equation}
s=(k_1+k_2)^2\,,\qquad  t=(k_1-k_3)^2\,,\qquad u=(k_1-k_4)^2\,,\qquad s+t+u=\sum_{i=1}^4 m_i^2\,.
\end{equation}
Since the polarizations for scalars are trivial, Eq.(\ref{eq:crossed2body}) implies that the amplitude for the crossed process 
$$\overline{\pi}^{\overline{a}_3}(k_1) \pi^{a_2}(k_2)\rightarrow \overline{\pi}^{\overline{a}_1}(k_3) \pi^{a_4}(k_4)$$
 can be obtained from $\mathcal{M}_{a_1 a_2 a_3 a_4}(s,t,u)$
  simply by sending $k_1\leftrightarrow -k_3$ and $\overline{a}_{1,3}\leftrightarrow a_{3,1}$, which corresponds to  the familiar crossing relation in terms of Mandelstam's variables
 \begin{equation}
 \mathcal{M}_{\overline{a}_3 a_2 \overline{a}_1 a_4}(s,t,u)=\mathcal{M}_{a_1 a_2 a_3 a_4}(u,t,s)\,.
 \end{equation}

For particles with spin the  polarizations are instead non-trivial, and exchanging $s\leftrightarrow u$ (and $\overline{a}_{1,3}\leftrightarrow a_{3,1}$) is not equivalent, in general, to crossing symmetry. See e.g. \cite{Bellazzini:2014waa,Espriu:2014jya}  for an explicit example where crossing massive spin-1 particles does not yield the same result of the transformation $s\leftrightarrow u$. Moreover, the amplitude is no longer an invariant scalar since it transforms under Lorentz transformations as the tensor product of the little-group (conjugate) representations $\mathcal{L}$ carried by the particles in the initial (final) state, i.e. 
\begin{equation}
\label{eq:lorentzT}
\mathrm{Lorentz}:\,\,|p^\sigma_a\rangle \rightarrow |(\Lambda p)^{\sigma^\prime}_a\rangle \, \mathcal{L}_{\sigma^\prime \sigma}(W(\Lambda,p))\,.
\end{equation}
For massive particles $\mathcal{L}$ is a $(2S+1)$-dimensional representation of the little group $SO(3)\sim SU(2)$, and $W(\Lambda,p)$ is the Wigner rotation.   The little group for massless particles is instead $ISO(2)$, but the translations inside $ISO(2)$ act trivially while the rotations give a phase, $\mathcal{L}_{\sigma^\prime \sigma}=\delta_{\sigma^\prime \sigma}e^{i\sigma\theta(W,p)}$. 

Nevertheless, the forward scattering where the kinematics of the initial state and the final state are the same, $k_i^{\sigma_i}=k_o^{\sigma_o}$, provides an exceptional configuration where special relations emerge.  The polarizations in (\ref{eq:crossed2bodyone}) combine into the form of density matrices (also known as spin projectors) of pure states
\begin{equation}
\label{eq:purestateden}
u^{\sigma}_\ell(\k) u^{\sigma\, \dagger}_{\ell^\prime} (\k) \equiv  \rho^{\sigma}_{\ell\ell^\prime}(\k)\,,\qquad v^{\sigma}_\ell(\k) v^{\sigma\, \dagger}_{\ell^\prime} (\k) \equiv \widetilde{\rho}^{\,\sigma}_{\ell\ell^\prime}(\k)
\end{equation}
(no sum on $\sigma$) which are traced with the amputated matrix elements, e.g. 
\begin{equation}
\mbox{forward:}\quad \mathcal{M}(k_{1\,a_1}^{\sigma_1}, k_{2\,a_2}^{\sigma_2}  \rightarrow  k_{1\,a_3}^{\sigma_1},  k_{2\,a_4}^{\sigma_2}  )= \rho^{\sigma_1}_{\ell_1 \ell^\prime_1}(\k_1)  \left[\mathcal{O}_{\ell^\prime_1 \ell_2^\prime}^{\ell_1  \ell_2}(k_1,k_2; a_1, a_2,a_3, a_4)\right] \rho^{\sigma_2}_{\ell_2 \ell^\prime_2}(\k_2)\,.
\end{equation}
 Under crossing of $1$ and $3$  one has   
\begin{equation}
\mbox{crossed-forward:}\quad \mathcal{M}(k_{1\,\overline{a}_3}^{\overline{\sigma}_1}, k_{2\,a_2}^{\sigma_2}    \rightarrow   k_{1\,\overline{a}_1}^{\overline{\sigma}_1}, k_{2\,a_4}^{\sigma_2} )= (-1)^{2S}\widetilde{\rho}^{\bar{\sigma}_1}_{\ell_1 \ell^\prime_1}(\k_1) \left[\mathcal{O}_{\ell^\prime_1 \ell_2^\prime}^{\ell_1  \ell_2}(-k_1,k_2; a_1, a_2,a_3, a_4)\right]\rho^{\sigma_2}_{\ell_2 \ell^\prime_2}(\k_2) \,.
\end{equation}
 The properties of these density matrices will allow us to extend them off-shell as  analytic functions of the 4-momentum (but not necessarily of the Mandelstam variables), and prove that the $s\leftrightarrow u$ exchange together with $\overline{a}_{1,3}\leftrightarrow a_{3,1}$ and $\sigma_1\leftrightarrow -\sigma_1$ is in fact equivalent, in the forward limit,  to the action of crossing.
  
  Moreover,  the use of a density matrix allows one to  generalize the analysis to actual mixed states which are described by density matrices $\rho_{\ell\ell^\prime}(\k)=\sum_\sigma p_\sigma  u^{\sigma}_\ell(\k) u^{\sigma\, \dagger}_{\ell^\prime} (\k)$ and $\widetilde{\rho}_{\ell\ell^\prime}(\k)=\sum_\sigma \tilde{p}_\sigma  v^{\sigma}_\ell(\k) v^{\sigma\, \dagger}_{\ell^\prime} (\k)$ where $p_{\sigma}$ and $\tilde{p}_{\sigma}$ are between 0 and 1. We discuss further this point in appendix~\ref{sec:sumrules}.

Hereafter we restrict to pure states density matrices (\ref{eq:purestateden}) and discuss in turn the massless and massive case.

\subsubsection*{Massless particles}

The little-group phases $\exp[i\sigma \theta(\Lambda,p)]$ that would arise from a Lorentz transformation in (\ref{eq:lorentzT}) actually cancel out between the initial and final state, in the forward scattering $t\rightarrow 0$, $\sigma_{1}=\sigma_{3}$, $\sigma_2=\sigma_4$. 
Hence, the forward amplitude for massless particles is in fact an invariant scalar where the helicity behaves just as an external label for the particles, on the same foot of the internal quantum numbers.

Consider a left-handed massless Weyl fermion: its pure state density matrix is a 2 by 2 spinor matrix which can be expressed in terms of the 4-momentum $p^\mu$ as
\begin{equation}
\label{rhopmW}
\rho^-(\p)=u^- (\p) u^{-\,\dagger} (\p)=v^+(\p) v^{+\,\dagger}(\p)=\widetilde{\rho}^{\,+}(\p)=p_\mu \sigma^\mu 
\end{equation}
where $p^0=|\p|$, $\sigma^{\mu}=(\mathbf{1},\sigma^i)$ and we used (\ref{eq:uvgamma5}). Expressed as a function of the 4-momentum, it is analytic and odd under $p^\mu\rightarrow -p^\mu$ compensating the sign change for crossing fermions, as well as the change in the polarizations  $u^+\leftrightarrow v^-$ and $u^{+\,\dagger}\leftrightarrow v^{-\,\dagger}$ for swapping particles/anti-particles of opposite helicities: 
\begin{equation}
\mbox{crossing massless spin-1/2:} \qquad \rho^{-}(p)\longrightarrow -\widetilde{\rho}^{\,+}(p)=\rho^{-}(-p)\,.
\end{equation}
 Therefore, crossing massless Weyl fermions in the forward limit acts on the whole amplitude, expressed in terms of the 4-momenta including also the the density matrices, simply as $k^{\sigma}_{a}\leftrightarrow -k^{-\sigma}_{\overline{a}}$ for the crossed particles
\begin{equation}
\label{crossingmzero2}
\mathcal{M}(-k^{-\sigma_1}_{1\,\overline{a}_3} k^{\sigma_2}_{2\, a_2}\rightarrow -k^{-\sigma_1}_{1\,\overline{a}_1} k^{\sigma_2}_{2\, a_4})=
\mathcal{M}(k^{\sigma_1}_{1\, a_1} k^{\sigma_2}_{2\, a_2}\rightarrow k^{\sigma_1}_{1\, a_3} k^{\sigma_2}_{2\, a_4})\,.
\end{equation}
Equivalently, it acts as $s\leftrightarrow -s$, $\sigma \leftrightarrow -\sigma$ and $a\leftrightarrow \overline{a}$ on the amplitude expressed with the Mandelstam variable $s$:
\begin{equation}
\label{crossingmzero}
\mathcal{M}(1^{-\sigma_1}_{\overline{a}_3} 2^{\sigma_2}_{a_2}\rightarrow 1^{-\sigma_1}_{\overline{a}_1} 2^{\sigma_2}_{a_4}, s)=
\mathcal{M}(1^{\sigma_1}_{a_1} 2^{\sigma_2}_{a_2}\rightarrow 1^{\sigma_1}_{a_3} 2^{\sigma_2}_{a_4}, -s)\,.
\end{equation}

The analytic properties of the density matrix for a massless spin-1 particle associated to a field $A_\mu$ that transforms like a Lorentz vector (up to gauge transformations) are fully analogous. Indeed, using Eq.~(\ref{eq:uvgamma5}),  the polarizations of opposite helicity of particles/anti-particles are in fact the same
\begin{equation}
\label{rhopmEps}
\rho^{\mp}_{\mu\nu}(\p)=\epsilon_{\mu}^{\mp} (\p) \epsilon_{\nu}^{\mp\,*} (\p)=\epsilon_{\mu}^{\pm\,*} (\p) \epsilon_{\nu}^{\pm} (\p)=\widetilde{\rho}^{\,\pm}_{\mu\nu}(\p) \,.
\end{equation}
Moreover,  up to irrelevant gauge transformations,  the polarizations are actually functions of the unit-vector $\hat{\p}=\p/|\p|$ through the rotation $R(\hat{\p})$ that sends the little-group reference vector $\k_r$ to $\p$, namely
$\epsilon^\sigma_\mu(\p)\equiv R_{\mu}^\nu(\hat{\p})e^\sigma_{\nu}(\k_r)$. The density matrix expressed as a function of the 4-momentum $p$ as $\rho^\sigma_{\mu\nu}(p)\equiv\rho^\sigma_{\mu\nu}(\hat{\p}=\p/p^0)$ is therefore invariant under $p\rightarrow -p$.  Equivalently, we can pick a convenient frame where $\p$ is aligned with $\k_r$ since the forward amplitude depends only on $s$. Hence, the action of crossing spin-1 massless particles in the forward limit is equivalent again to $k^{\sigma}_{a}\leftrightarrow -k^{-\sigma}_{\overline{a}}$ everywhere including the contribution from the polarizations up to gauge transformations,
\begin{equation}
\mbox{crossing massless spin-1:} \qquad \rho^{\pm}_{\mu}(p)\longrightarrow \widetilde{\rho}^{\mp}_{\mu\nu}(p)=\rho^{\pm}_{\mu\nu}(-p)\,,
\end{equation}
implying the relations~(\ref{crossingmzero2}) and (\ref{crossingmzero}). 

This result trivially extends to massless gravitons since their polarizations are $\epsilon^{\pm}_{\mu\nu}(\p)=\epsilon^{\pm}_{\mu}(\p) \epsilon^{\pm}_{\nu}(\p)$, or e.g. massless spin-3/2 with $u^{\pm}_{\mu\,\alpha}=\epsilon^{\pm}_\mu u^{\pm}_\alpha$. 
In fact, it extends to higher spins associated to massless fields that can be chosen transforming  as covariant representations $(S,0)$ or $(0,S)$ of the Lorentz group $SO(3,1)$ such as a self-dual field strength $F_{\mu\nu}$ in the $(1,0)$ or $(0,1)$ representation. Indeed, one can show \cite{Weinberg:1964ev,Weinberg:1969di} that the density matrices can be expressed as monomials in the 4-momentum $p_\mu$ of order $2S$. Together with $CPT$ invariance this implies 
\begin{equation}
\mbox{crossing massless spin-S:}\qquad  \rho^{\pm}(p)\longrightarrow (-1)^{2S}\widetilde{\rho}^{\mp}(p)=\rho^{\pm}(-p) 
\end{equation}
 and hence the crossing relations~(\ref{crossingmzero2}) and (\ref{crossingmzero}).
These relations for the analytically continued density matrices, $\rho^{\sigma}(k)=\widetilde{\rho}^{\,-\sigma}(k)=(-1)^{2S}\rho^\sigma(-k)$, are such to enforce locality of the free theory i.e.  
\begin{equation}
\label{eq:loc}
[\Psi_{\ell_1}(x_1),\Psi^\dagger(x_2)_{\ell_2}]_{\pm}=\rho_{\ell_1\ell_2}(i\partial)\int \frac{d^3k}{2|k|(2\pi)^3}\left( e^{-ik(x_1-x_2)} \pm (-1)^{2S}e^{ik(x_1-x_2)}\right)\big|_{(x_1-x_2)^2<0}=0
\end{equation}
at space-like distances, where $[\,,]_{\pm}$ is the commutator ($-$) or the anti-commutator $(+)$.    This implies locality of the interacting theory (e.g. via K\"all\'en-Lehmann decomposition \cite{Weinberg:1995mt}) and hence causality \cite{Dubovsky:2007ac}. 
Since the density matrices are nothing but the numerators of free propagators, i.e. 
\begin{equation}
\langle\mathrm{T}\Psi_{\ell_1}(x_1)\Psi^\dagger_{\ell_2}(x_2)\rangle = \int  \frac{d^4k}{(2\pi)^4} e^{-ik(x_1-x_2)} \frac{i\,\rho_{\ell_1\ell_2}(k)}{k^2-i\epsilon}\,,
\end{equation}
they enforce as well the correct commutation or anti-commutation relations are required by the spin-statistics theorem for the propagators continued analytically to the euclidean signature.

\subsubsection*{Massive particles}
\label{subsec:massivecrossing}

Whenever a massive particle is involved in the scattering process, the forward amplitude is no longer Lorentz invariant but transforms according to Eq.~(\ref{eq:lorentzT}).
This simply requires us to specify a reference frame: we pick the centre of mass frame and orient the $z-$axis, that is the direction where the spins are measured, along the momentum of the incoming particle $1$, $k_1=(k_1^0,0,0,k^z_1)^T$. We are dealing i.e. with helicity amplitudes, even though helicity itself is not Lorentz invariant.
Boosts and rotations along the $z$-axis transform the states but leave the forward amplitude invariant. Therefore,  it must be a function of the only non-vanishing invariant under the 2D Lorentz group in this special kinematics, that is the Mandelstam variable $s$.

Let us start with a Dirac or Majorana fermion $\Psi$ of mass $m_\Psi$: the density matrix is a 4 by 4 matrix with spinor indexes that can be expressed as
\begin{equation}
\label{densfermion1}
\rho^{\sigma} = u^{\sigma}(\k_1)u^{\sigma\,\dagger}(\k_1)=(\slashed{k_1} + m_\Psi)\frac{1+\gamma^5\slashed{a}^\sigma(\k_1)}{2}\gamma^0
\end{equation}
where $a_\mu^\sigma$ is the polarization 4-vector, see appendix~\ref{app:polarizations} and e.g. \cite{Berestetsky:1982aq}. In our kinematics it takes the form
\begin{equation}
\label{eq:aoffshell}
a_\mu^{\pm}(\k_1)=\pm\frac{1}{m_\Psi}\left(k^z_1, 0, 0, k^0_1\right)^T   
\end{equation}
obtained by boosting the rest-frame's polarization vector $a^{\pm}(\mathbf{0})=(0,0,0,\pm 1)^T$ along the $z-$direction with velocity $\beta=-k^z/k^0$, that is applying the Lorentz transformation $\Lambda_\mu^\nu(k_1)$ 
\begin{equation}
\label{boostrestcom}
\Lambda_{\mu}^\nu(k_1)=\left(
\begin{array}{cccc}
  \dfrac{k_1^0}{m_\Psi}   &  0  &   0 & \dfrac{k_1^z}{m_\Psi}  \\
  0           &  1  &  0 & 0  \\
  0           &  0  &  1 & 0\\
  \dfrac{k_1^z}{m_\Psi}   &  0  &  0 & \dfrac{k_1^0}{m_\Psi} 
\end{array}
\right)\,,
\end{equation}
which is linear in the 4-momentum $k_1$. This allows us to analytically continue the polarization vector in (\ref{eq:aoffshell}) to a function of the 4-momentum which is linear in $k$ too, 
\begin{equation}
a_\mu^{\pm}(\k)\rightarrow a_\mu^{\pm}(k)= -a_\mu^{\pm}(-k)\,.
\end{equation}

For an anti-particle moving in the same direction $\k_1$, using again Eq.~(\ref{eq:uvgamma5}), we have
\begin{equation}
\label{densfermion2}
\widetilde{\rho}^{\,\sigma} = v^{\sigma}(\k_1)v^{\sigma\,\dagger}(\k_1)=(\slashed{k_1} - m_\Psi)\frac{1+\gamma^5\slashed{\tilde{a}}^{\sigma}}{2}\gamma^0 \qquad \tilde{a}^\sigma_\mu=-a^{-\sigma}_\mu\,.
\end{equation}
Note that for $m_\Psi\rightarrow 0$ the polarization vector reduces to $a_\mu^{\pm}(k)\rightarrow \pm k^\mu/m_{\Psi}$ and one smoothly recovers the massless relation (\ref{rhopmW}) written in the 4-component notation, $\rho^{\pm}(k)=P_{\pm}\slashed{k}\gamma^0= \widetilde{\rho}^{\,\mp}(k)$, where $P_{\pm}=(1\pm\gamma^5)/2$ are the projectors over the right and left chiralities respectively.

Expressed in terms of the 4-momentum, the relation between particle/anti-particle density matrix for spin-1/2 is $\widetilde{\rho}^{\,\mp}(k_1)=-\rho^{\pm}(-k_1)$, implying once again Eq.~(\ref{crossingmzero2}) and
\begin{equation}
\label{crossingmassivespin}
\mathcal{M}(1^{-\sigma_1}_{\overline{a}_3} 2^{\sigma_2}_{a_2}\rightarrow 1^{-\sigma_1}_{\overline{a}_1} 2^{\sigma_2}_{a_4}, s)=
\mathcal{M}(1^{\sigma_1}_{a_1} 2^{\sigma_2}_{a_2}\rightarrow 1^{\sigma_1}_{a_3} 2^{\sigma_2}_{a_4}, u)\,
\end{equation}
where $t=0$ and $u=-s+\sum_i m_i^2$.

The polarizations of spin-1/2 particles contain, for a finite mass, non-analyticities in the Mandelstam variables that may or may not propagate to the whole amplitude. For example, in the $\Psi \psi\rightarrow \Psi \psi$ scattering mediated by the $P$- and $C-$violating interaction $\bar{\Psi}\gamma^\mu\gamma^5\Psi \bar{\psi}\gamma_\mu\psi$ (where we assume for simplicity $m_\psi=m_\Psi$), the forward amplitude is proportional to $m_\Psi a^{\sigma_1}_\mu(\k_1) k_2^\mu \propto \sqrt{-us}$ which has a tree-level branch-cuts in the complex $s$-plane from $s=0$ to $s=4m_\Psi^2$. We discuss in section~\ref{sec:unitarity} the role of this discontinuity on the dispersion relations and the resulting positivity constraints.

Passing to massive spin-1 bosons in the vector representation, we can work directly with an explicit basis obtained by boosting along the $z-$axis the eigenvectors of $S_z$ in the particle rest frame 
\begin{equation}
\epsilon^{\pm}_\mu(\k_1)=\mp \frac{1}{\sqrt{2}}(0,1,\pm i,0)^T\,,\qquad \epsilon^0_{\mu}(\k_1)=\frac{1}{m_\Psi}(k^3_1,0,0,k^0_1)^T\,.
\end{equation}
Clearly, the particle/anti-particle density matrices expressed in terms of the 4-momentum are related by $\widetilde{\rho}^{\sigma}(k)=\rho^{-\sigma}(k)=\rho^{-\sigma}(-k)$. Therefore, the crossed scattering amplitudes are again related as in Eq.~(\ref{crossingmzero2}) and (\ref{crossingmassivespin}) simply by $s\leftrightarrow u$, and $\overline{a}\rightarrow a$ for the internal quantum numbers.

These results extend to massive higher spins by taking suitable tensor products between the  polarizations that we have studied so far. Consider for example a tensor $\Psi_{\mu_1\ldots \mu_n}$ that destroys a particle with integer spin as in (\ref{overlap}). Its polarizations $u_\ell(\mathbf{0})=u_{\mu_1\ldots \mu_n}(\mathbf{0})$ in the rest frame are boosted along the $z$-axis by acting on each index with $\Lambda_\mu^\nu(k_1)$ of Eq.~(\ref{boostrestcom}), which is linear in the 4-momentum $k_1$, 
resulting again in $\widetilde{\rho}^{\,\pm}(k)=\rho^{\mp}(k)=\rho^{\mp}(-k)$ and hence Eq.~(\ref{crossingmassivespin}). Analogously, higher half-integer spins can be found in the tensor product of vector and spin-1/2 representations.  As example consider a massive spin-3/2 with polarization $u^\sigma_{\mu\,\alpha}=\sum_{\sigma^\prime\sigma^{\prime\prime}}C^{\sigma}_{\sigma^\prime \sigma^{\prime\prime}}\epsilon^{\sigma^\prime}_\mu \cdot u^{\sigma^{\prime\prime}}_\alpha $ where $C^{\sigma}_{\sigma^\prime \sigma^{\prime\prime}}$ is the Clebsch-Gordan coefficient $\langle(1/2, \sigma^{\prime\prime}) (1,\sigma^\prime)| (3/2,\sigma) \rangle$.

\section{Unitarity constraints and positivity bounds} \label{sec:unitarity}

\begin{figure}[htb]
\centering
\includegraphics[width=10cm]{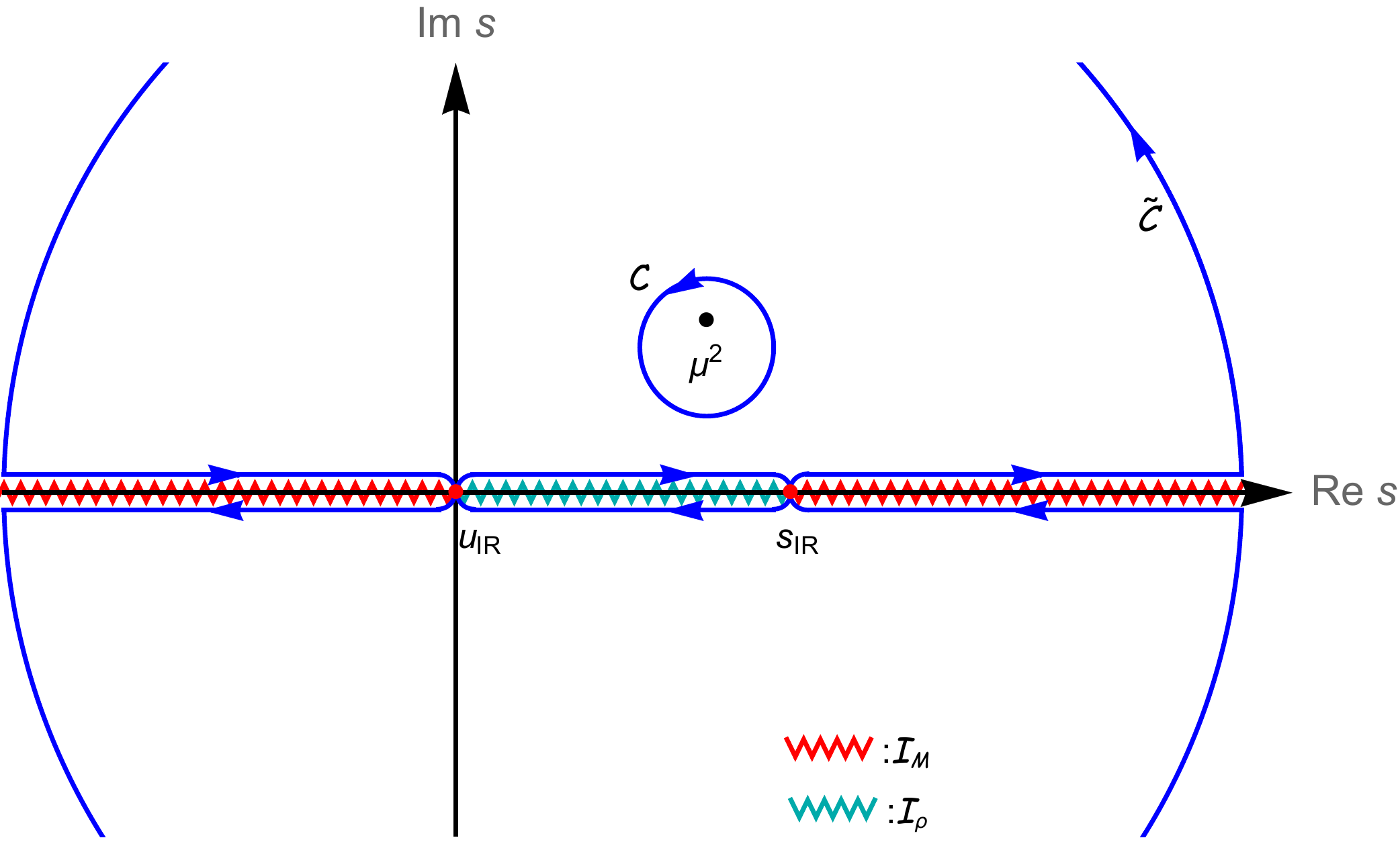}
\caption{Contours in the complex $s$-plane in the forward elastic scattering with $t=0$. The branch-cut of the square-root type represented by a cyan saw-like line in the interval $\mathcal{I}_\rho=(u_{IR},s_{IR})$ may arise only from the polarizations of massive half-integer spins, and only for certain parity violating interactions, see main text and Eq.~(\ref{Eq:examplediscfer}). The standard discontinuities of the scattering amplitudes are represented by a red saw-like lines $\mathcal{I}_{\mathcal{M}}$; they come from the amputated correlation functions. At the branch-points $s_{IR}=(m_1+m_2)^2$ and $u_{IR}=(m_1-m_2)^2$, associated with the 2-particle elastic thresholds, the amplitude and the discontinuities vanish.}
\label{fig:contourspin}
\end{figure}

We consider now forward and \textit{elastic} amplitudes where not only the kinematical variables $k^\sigma$ but even the internal quantum numbers $a$  are the same in the $in$ and $out$ states, namely $a_1=a_3$ and $a_2=a_4$ together with $k_1^{\sigma_1}=k_3^{\sigma_3}$ and $k_2^{\sigma_2}=k_4^{\sigma_4}$. Displaying fewer indexes for convenience,
$$ \mathcal{M}(1^{\sigma_1}_{a_1} 2^{\sigma_2}_{a_2}\rightarrow 1^{\sigma_1}_{a_1} 2^{\sigma_2}_{a_2}; \,s)\equiv \mathcal{M}_{a_1a_2}^{\sigma_1 \sigma_2}(s)\,,$$
we expand the amplitude around a point $s=\mu^2$ in the complex $s$-plane where it is analytic
\begin{equation}
\mathcal{M}_{a_1a_2}^{\sigma_1 \sigma_2}(s)= \mathcal{M}_{a_1a_2}^{\sigma_1 \sigma_2}(\mu^2)+ (s-\mu^2)\, \mathcal{M}_{a_1a_2}^{\prime\,\sigma_1 \sigma_2}(\mu^2)+\frac{1}{2!}(s-\mu^2)^2\,\mathcal{M}_{a_1a_2}^{\prime\prime\,\sigma_1 \sigma_2}(\mu^2)+\ldots 
\end{equation}
The  primes ${}^\prime$'s represent derivatives with respect to $s$. The Taylor coefficients can be extracted with the Cauchy integral formula, e.g. 
\begin{equation}
\label{eq:seconderivative}
\mathcal{M}_{a_1a_2}^{\prime\prime\,\sigma_1 \sigma_2}(\mu^2)=\frac{2!}{2\pi i} \oint_{\mathcal{C}} \frac{ds}{(s-\mu^2)^3} \mathcal{M}_{a_1a_2}^{\sigma_1 \sigma_2}(s)\,,
\end{equation}
where $\mathcal{C}$ is \textit{any} contour in the complex $s-$plane that encloses $s=\mu^2$ but no other singularity, see Fig.~\ref{fig:contourspin}.  
Choosing $\mu^2$ in the IR,  the taylor coefficients on the left-hand side of (\ref{eq:seconderivative}) can be expressed in terms of the Wilson coefficients $c_i$ of the effective lagrangian 
\begin{equation}
\label{eq:EFTL}
\mathcal{L}_{EFT}=\sum_i c_i\, \mathcal{O}_i\,,
\end{equation}
that describes the dynamics of the IR degrees of freedom at low-energy. The Wilson coefficients and the EFT itself are indeed designed to match the full amplitude when evaluated in the IR.   
On the right-hand side of (\ref{eq:seconderivative}), the contour can be deformed into $\widetilde{\mathcal{C}}$  running over the branch-cuts (and possibly poles on the real axis if any) and a big circle eventually sent to infinity, see Fig.~\ref{fig:contourspin}. The variable $s$  under the integral along $\widetilde{\mathcal{C}}$ may take very large values, well above the cutoff of the effective theory. Needless to say, one should not use the EFT lagrangian (\ref{eq:EFTL}) to evaluate the right-hand side of Eq.~(\ref{eq:seconderivative}). In fact, we do not want to, and we would not be able to,  calculate the contour integral. It would be a herculean task that would require the knowledge of the underlying theory up to arbitrary high-energy.  We rather want to show that the integral is positive in any underlying unitary theory that UV completes the effective lagrangian, yielding in turn a positivity constraints on the low-energy Wilson coefficients via Eq.~(\ref{eq:seconderivative}) and (\ref{eq:EFTL}). To this end we need to understand the analytic structure of the amplitude.

For massive particles, the scalar functions in the amputated matrix elements, such as e.g. the functions $a_i(s)$  in the fermion-scalar scattering
\begin{equation}
\label{scalfunctpiN}
\mathcal{M}=\Tr \left\{\rho(\k_1) \left[a_s(s)+ a_A(s)\Gamma^5 + a_V(s)\gamma^\mu k_{2\,\mu}+ a_{PV}(s)\gamma^5 \gamma^\mu k_{2\,\mu} \right]\right\}\,,
\end{equation}
 are analytic on the real axis below the thresholds for the $s$-channel and $u$-channel branch-cuts, except possibly for isolated poles of light particles that can be exchanged in the scattering. As long as the contraction with the polarizations, i.e. the trace with the density matrix $\rho(\k_1)$, does not introduce further branch-cuts below those thresholds, the full elastic forward amplitude can be extended to a real function of the complex cut-plane of $s$ variable   \cite{OliveSmatrix,Olive:1962}
\begin{equation}
\label{eq:realcond}
\mathcal{M}_{a_1a_2}^{\sigma_1 \sigma_2\, *}(s)=\mathcal{M}_{a_1a_2}^{\sigma_1 \sigma_2}(s^*)
\end{equation}
via the Schwarz reflection principle.  This relation is important because it allows us to link the discontinuity across the cuts to the imaginary part of the elastic forward amplitude, and eventually to the total cross-sections via the optical theorem\footnote{We follow the conventions of \cite{Schwartz:2013pla} where the scattering matrix  is $S=1+(2\pi)^4\delta^4(\sum p_i)i\mathcal{M}$. We work with the mostly minus' signature $(+,-,-,-)$ of spacetime.}
\begin{equation}
\label{Eq:opticalth}
\mathrm{Im}\mathcal{M}^{\sigma_1\sigma_2}_{a_1 a_2}(s+i\epsilon)=  \sqrt{(s-m_1^2-m_2^2)^2-4m_1^2m_2^2}\times
 \sigma^{\mathrm{tot}}(1_{a}^{\sigma_1}  2_{a_2}^{\sigma_2} \rightarrow \mathrm{anything})(s)
 \end{equation}
 that follows from unitarity of the $S$-matrix for $\epsilon\rightarrow 0^{+}$ and $s\geq (m_1+m_2)^2$. Of course, inequalities are always understood with the restriction to $s\in \mathbb{R}$.

The analytic structure we described so far represents the common situation for integer spins. For example, scattering massive spin-1 particles the only discontinuity that could possibly arise from the polarizations  would come from the longitudinal polarizations via a term $\epsilon^{\sigma=0}_\mu(\k_1) k_2^\mu\propto \sqrt{-su}$ which, however,  can only appear squared in the full amplitude (see appendix~\ref{app:polarizations}). Therefore, it does not change the analytic structure of the full amplitude which still respects the reality condition of Eq.~(\ref{eq:realcond}).

The story for fermions is slightly more complicated but, nonetheless,  Eq.~(\ref{eq:realcond})  still holds true, as we discuss below. We have seen in the previous section that certain $P$-violating interactions that involve half-integer spins can give rise to non-analyticities in the density matrix which may in fact be transmitted to the full  amplitude through the polarization vectors $a^\sigma_\mu(\k)$ of Eq.~(\ref{densfermion1}), e.g.
\begin{equation}
\label{Eq:examplediscfer}
\mathcal{M} \supset a^{\pm}_\mu(\k_1) k_{2}^\mu  =\pm \frac{1}{2m} \sqrt{-su+(m_1^2-m_2^2)^2} \,.
\end{equation}
A 4-fermion interaction $\bar{\Psi}\gamma^\mu\gamma^5\Psi \bar{\psi}\gamma_\mu\psi$ provides an example that gives rise to such a non-analytic behavior due to the polarizations of $\Psi$. 
Another example from scattering spin-1/2 fermion off a longitudinally polarised spin-1 boson is $a^{\pm}_\mu(\k_1) \epsilon^{0\,\mu}(\k_{2})$ which is proportional again to the same square-root (\ref{Eq:examplediscfer}).

 Nevertheless, these discontinuities from the density matrices are of the square-root type and have the branch-cuts of finite support on the real axis in the interval $\mathcal{I}_{\rho}=(u_{IR},s_{IR})$, where $s_{IR}=(m_1+m_2)^2$ and $u_{IR}=(m_1-m_2)^2$. That is, the density matrices are continuous and real for larger values of $|s|$, and can thus be extended to the whole cut-plane $\mathbb{C}\setminus \mathcal{I}_{\rho}$. Analogously, the scalar functions in the amputated matrix elements are non-analytic in a (complementary) region $\mathcal{I}_{\mathcal{M}}$, but they can be analytically extended to real functions of the cut plane $\mathbb{C}\setminus \mathcal{I}_{\mathcal{M}}$, see Fig.~\ref{fig:contourspin}. Therefore, the full amplitudes for fermions still satisfy the reality condition~(\ref{eq:realcond}) in $\mathbb{C}\setminus (\mathcal{I}_{\rho} \cup \mathcal{I}_{\mathcal{M}})$. Note that the discontinuities at the branch-points associated to the thresholds $s_{IR}$ and $u_{IR}$ vanish. 

 We stress once more that that this branch-cut $\mathcal{I}_\rho$, whenever present, comes entirely from the density matrices, that is from the external polarizations. It is not there for the scalar functions $a_i(s)$ of the type (\ref{scalfunctpiN}) that are usually considered, e.g., in $\pi-N$ scattering \cite{Hamilton:1963zz}.  Since we are interested in scattering arbitrary spins, it would be very inefficient, if possible at all,  to work with the analog of those  scalar amplitudes $a_i(s)$ since one would need to perform a Lorentz decomposition for any form factor for generic spins.  The positivity conditions (\ref{eq:positivity1}) and (\ref{eq:positivity2}) show that splitting the amplitudes in the scalar functions is neither needed nor useful in general: it suffices to work with the actual amplitudes whose discontinuity are also readily expressed in terms of total cross-sections. Besides, in parity preserving theories like QCD these discontinuities along $\mathcal{I}_\rho$ are not generated  anyway.

With the reality condition at hand, we can identify the imaginary part of the amplitude with its discontinuity along the real axis
\begin{equation}
2i\,\mathrm{Im}\mathcal{M}^{\sigma_1\sigma_2}_{a_1 a_2}(s+i\epsilon)=\left[\mathcal{M}^{\sigma_1\sigma_2}_{a_1 a_2}(s+i\epsilon)-\mathcal{M}^{\sigma_1\sigma_2}_{a_1 a_2}(s-i\epsilon)\right]\qquad s\in\mathbb{R}\,,
\end{equation} 
and we can thus split the contour along the branch-cuts into three integrals over the  imaginary parts, 
\begin{align}
\label{eq:dispn}
\mathcal{M}_{a_1a_2}^{\prime\prime\,\sigma_1 \sigma_2}(\mu^2) = & \frac{2!}{\pi}\left(\int^\infty_{s_{IR}} \frac{ds}{(s-\mu^2)^3} +\int_{-\infty}^{u_{IR}} \frac{ds}{(s-\mu^2)^3}\right) \mathrm{Im}\mathcal{M}^{\sigma_1\sigma_2}_{a_1 a_2}(s+i\epsilon)\\
\nonumber
& +   \frac{2!}{\pi}\int_{u_{IR}}^{s_{IR}} \frac{ds}{(s-\mu^2)^3}\mathrm{Im}\mathcal{M}^{\sigma_1\sigma_2}_{a_1 a_2}(s+i\epsilon)+C_{\infty}\,,
\end{align}
and the integral $C_\infty$ along the big circle whose radius is eventually sent to infinity. Assuming that the asymptotic amplitude grows less than $s^2$ for $s\rightarrow \infty$, which is always the case for gapped theories thanks to the Froissart bound \cite{Froissart:1961ux}, we can drop the big circle's contribution which vanishes at infinity, 
\begin{equation}
C_{\infty}\rightarrow 0\,.  
\end{equation}
Moreover, by changing variables $s\rightarrow u=-s+2(m_1^2+m_2^2)$ and using the crossing relation~(\ref{crossingmassivespin}) that we have proven for any spin, we can recast the integral over the $u$-channel as an integral over the physical energies of the $s$-channel scattering where particles $1$ and $3$ have been replaced by their antiparticles of opposite spins\footnote{For self-conjugate particles one may prefer working with linear polarizations; in that case the index $\sigma$ labelling the linear polarizations would not be flipped under crossing, see Appendix~\ref{sec:crossforw}.} and internal quantum numbers, namely 
\begin{equation}
\int_{-\infty}^{u_{IR}} \frac{ds}{(s-\mu^2)^3}\mathrm{Im}\mathcal{M}^{\sigma_1\sigma_2}_{a_1 a_2}(s+i\epsilon)=\int_{s_{IR}}^{\infty} \frac{ds}{(s-2(m_1^2+m_2^2)+\mu^2)^3} \mathrm{Im}\mathcal{M}^{-\sigma_1\sigma_2}_{\overline{a}_1 a_2}(s+i\epsilon)\,.
\end{equation}

So far we have used only analyticity and crossing symmetry of the scattering amplitude.  Requiring the unitarity of the underlying UV theory, the optical theorem (\ref{Eq:opticalth}) implies positive imaginary parts  above thresholds $s\geq s_{IR}$, namely
\begin{align}
\mathrm{Im}\mathcal{M}^{\sigma_1\sigma_2}_{a_1 a_2}(s+i\epsilon)  \geq 0\,,\qquad \mathrm{Im}\mathcal{M}^{-\sigma_1\sigma_2}_{\overline{a}_1 a_2}(s+i\epsilon)  \geq 0\,,
\end{align}
The inequality is saturated only for the trivial theory where particle $1$ and $2$ (and $\bar{1}$ and 2) do not interact with each other so that $\sigma^{\mathrm{tot}}(1_{a}^{\sigma_1}  2_{a_2}^{\sigma_2} \rightarrow \mathrm{anything})=0$ and $\sigma^{\mathrm{tot}}(\bar{1}_{\bar{a}}^{-\sigma_1}  2_{a_2}^{\sigma_2} \rightarrow \mathrm{anything})=0$.
Analyticity, crossing symmetry and unitarity imply thus the following positivity constraint on the IR scattering amplitude for an interacting theory
\begin{equation}
\label{eq:positivity1}
\mathcal{M}_{a_1a_2}^{\prime\prime\,\sigma_1 \sigma_2}(\mu^2)- \frac{2!}{\pi}\int_{u_{IR}}^{s_{IR}} \frac{ds}{(s-\mu^2)^3}\mathrm{Im}\mathcal{M}^{\sigma_1\sigma_2}_{a_1 a_2}(s+i\epsilon) > 0
\end{equation}
as long as $\mu^2$ is sent to the real axis from above and lies between $(m_1\pm m_2)^2$, for example at the crossing symmetric point $\mu_c^2=m_1^2+m_2^2$. 
This relation simply states that the contour integral encircling $\mu^2$ and the IR branch-cut $\mathcal{I}_\rho$ from the density matrix (if any) is positive.~\footnote{We have omitted so far, just for easy of presentation, the residues of other IR poles on the real axis between $s_{IR}$ and $u_{IR}$ that would have appeared subtracted on the left-hand side of Eq.~(\ref{eq:positivity1}) just like the contribution from the IR branch-cut. They can be in fact shuffled inside that integral around the interval $\mathcal{I}_{\rho}$. Analogously, the UV poles along $\mathcal{I}_{\mathcal{M}}$ can be shuffled inside the integral along $\mathcal{I}_{\mathcal{M}}$; they just add another positive contribution to the right-hand side of (\ref{eq:positivity1}). Eq.~(\ref{eq:positivity1}) simply states that the contour integral encircling $\mu^2$, the IR branch-cut $\mathcal{I}_\rho$, and the IR poles is positive.} 
 The most important point is that both $\mu^2$ and $\mathcal{I}_{\rho}$ are in the IR and we can thus evaluate the left-hand side of (\ref{eq:positivity1}) with the $EFT$ Lagrangian (\ref{eq:EFTL}). In turn, this positivity condition enforces inequalities on the EFT Wilson coefficients. 

As we have already remarked previously, the IR branch cut is often absent. This happens  e.g. in any theory with only integer spins, or in parity preserving theories, or in the massless limit (more on this limit below).  The dispersion relation reduces in such cases to the neat expression~\footnote{Such a simple and neat expression actually holds in full generality, i.e. even in the occurrence of the IR branch cuts for massive particles, provided one is summing over the polarizations of the half-integer spins, e.g. \begin{equation*}\sum_{\sigma_1, \sigma_2} \mathcal{M}_{a_1a_2}^{\prime\prime\,\sigma_1 \sigma_2}(\mu^2) >0\,, \end{equation*}  since the averaged density matrix is an analytic function (in fact, it follows by causality (\ref{eq:loc}) that it is a polynomial of order $2S$) containing no IR branch-cut $\mathcal{I}_\rho$. Since this is mathematically equivalent to consider fully un-polarized spinning particles, it is no surprise that one recovers the inequality that holds for scalars.}
\begin{equation}
\label{eq:positivity2}
\mathcal{M}_{a_1a_2}^{\prime\prime\,\sigma_1 \sigma_2}(\mu^2) > 0\,,
\end{equation}
where $(m_1-m_2)^2 \leq \mu^2\leq (m_1+m_2)^2$.

We stress that these inequalities are exact results that hold non-perturbatively for particles of arbitrary spin as they are derived from first principles of the S-matrix theory.~\footnote{Should one work at tree-level, Eq.~(\ref{eq:positivity1}) could be turned into a neat expression similar to (\ref{eq:positivity2}), even when IR branch-cut from the polarizations or other IR poles below threshold are present, thanks to a simple trick from complex analysis. Because the tree-level EFT has no UV branch cut, the left-hand side in (\ref{eq:positivity1}) is nothing but (minus) the residue at infinity calculated with the tree-level EFT: \begin{equation*}\mathcal{M}_{a_1a_2}^{\prime\prime\,\sigma_1 \sigma_2}(m_i^2\ll \mu^2 \ll \Lambda^2 )\big|_{EFT,tree} > 0\,.\end{equation*}  In practice, the low-energy integral along the branch-cut $\mathcal{I}_\rho$ produced by the polarizations on the right-hand side of (\ref{eq:positivity1}) appears just to remove the sensitivity to the choice of the IR data, such as  the particles' masses $m_i^2$,  or any other IR scale. Of course, one should never restrict to such a tree-level argument for EFTs that at tree-level give amplitudes of $O(s)$. In such a case one should perform loops to properly evaluate the contribution on the left-hand side of the inequality (\ref{eq:positivity1}), as e.g. done in  \cite{Mateu:2008gv} for the chiral lagrangian in QCD.}

Note that we could take even further subtractions, that is obtain positivity conditions for higher derivatives of the amplitude, as long as the gap is open  and $\mu^2\neq 0$ in order to regulate possible IR divergences. 

\subsubsection*{Massless limit}
So far we assumed the spectrum was gapped, but what about a theory with only massless particles? We can deform it by adding at least one mass $m$ to regulate the IR, open the gap, infer the reality condition, derive then the positivity bound (\ref{eq:positivity2}) for arbitrarily small $m$, and eventually take the limit $m\rightarrow 0$ while retaining the positivity of the IR Wilson coefficients.  There could be though three points that could spoil this program.

\begin{itemize}
 \item First, adding a mass may require to add extra IR degrees of freedom. This happens e.g. for the theory of massive gauge bosons and massive gravity where the gauge bosons and the graviton eat Goldstone bosons and galileons modes, respectively. One should carefully identify which degrees of freedom produce the leading $s^2$ growth of the amplitude before taking the zero mass limit. Should the $s^2$ behavior be generated instead by the extra degrees of freedom, one would obtain the positivity bound for those modes rather than for the ones of interest. 
 
\item Second, the forward limit $t\rightarrow 0$, the massless limit $m_i\rightarrow 0$,  and sending the big circle to infinity may be singular limits and/or not commute with each other when massless particles of spin $S\geq1$ are exchanged in the t-channel.  Extra assumptions may be required in this  case~\footnote{We should also stress that, when scattering massless spins higher or equal than 1, the degree of the polynomial bounding the elastic forward amplitude could be higher or equal than $s^2$ \cite{Adams:2006sv}; in such a case one should trivially take the smallest number of even derivatives that allow to discard the contribution to the dispersion relation that comes from the big circle at infinity. Alternatively, one can work at tree-level only, as e.g. in \cite{Adams:2006sv,Bellazzini:2015cra}. }, see e.g. \cite{Bellazzini:2014waa,Bellazzini:2015cra,Nicolis:2009qm}. 
 Alternatively, one may simply study the limit where these integer-spin massless modes in t-channel decouple while assuming that the resulting theory remains consistent. This is e.g. what one implicitly does when studying a non-gravitational theory and neglects gravity, i.e. he/she takes the limit $M_{Planck}\rightarrow\infty$. Analogously for the photon, one assumes that turning off the weak gauging of a global  $U(1)$ symmetry does not back-react so strongly on the system under study to destroy its consistency. 
 
 \item Third, the positivity bound (\ref{eq:positivity2}) on the leading $s^2$-terms of the scattering matrix of massless particles is valid only for sufficiently soft amplitudes, in order to ensure the IR convergence in the limit $m_i\rightarrow 0$. For example, the interaction $g_*^2(\bar{\psi}\psi)^2/\Lambda^2$ for massless fermions is not soft enough as it gives a total cross-section below any other threshold $\sigma^{tot}(s\rightarrow 0)=\sigma^{elastic}\sim s \cdot g_*^4/\Lambda^4$ and thus $\mathrm{Im}\mathcal{M}(s\rightarrow 0)\sim s^2\cdot g_*^4/\Lambda^4$. This behavior is not enough to grant the IR convergence of the dispersion relation for $\mathcal{M}^{\prime\prime}$. Equivalently, the second derivative of the amplitude is going to be IR divergent in $s=t=0$ because of the dispersive integral that relates cross-sections and amplitudes.  
 In this case only a once-subtracted dispersion relation for $\mathcal{M}^\prime$ would be IR convergent; one would need then to make extra assumptions about the high energy behavior of the amplitude as in \cite{Bellazzini:2014waa,Low:2009di}, since the Froissart bound is not longer enough to discard the big circle's contribution $C_{\infty}$ with just one subtraction.  Moreover, the resulting expression would not necessarily imply a positivity constraint as the $u$- and $s$-channel contribution enters with opposite signs under the dispersive integral for an odd-number of subtractions. We come back to once-subtracted sum rules for dimension-6 operators in appendix~\ref{sec:sumrules}. 
 \end{itemize}

 Interestingly, neither the first nor the second problem described above arise for scattering massless scalars and/or massless spin-1/2 fermions. The third point about the IR convergence simply requires amplitudes as soft as $s^{n}$, or finite masses, in order to place a dispersion relation on the $n$-th derivative $\mathcal{M}^{(n)}$. In the next section we study examples of such a soft theories.  
 Whether the amplitudes are sufficiently well behaving in the IR to admit a massless limit can be established by direct inspection of the EFT at hand which, by construction, reproduces the correct IR behavior of the underlying fundamental theory.  Should such a good IR behavior not  be granted, one would need to work with one or more massive states.

\section{Soft limits}
\label{sec:softlimi}

As it was stressed e.g. in \cite{Cheung:2014dqa,Cheung:2015ota}, it is often possible to reconstruct a theory and its symmetries by the leading IR behavior of the scattering amplitudes: the softer the amplitude the more symmetry is required in order to cancel the would-be leading terms. Notorious examples include gauge and gravity theories, non-linear sigma models, dilatons, and galileons \cite{Nicolis:2008in}. It is thus relevant to ask how soft the scattering amplitudes for fermions can be, and which symmetries are associated with the enhanced soft-behavior of the amplitudes.  
Moreover, particles that are highly boosted but with energy below the cutoff  $\Lambda$ can be considered both massless and soft compared to $\Lambda$ . The leading soft behavior tells us how fast the amplitude can raise with energy in such highly  boosted regime but within the validity of the EFT.

\subsection{Soft limits for spin-1/2}

We take a step in this direction by studying the low-energy theory of a massless spin-1/2 field $\chi$.  The mass term can be forbidden e.g. by chiral symmetry transformations 
\begin{equation}
\label{eq:chiralsym}
\chi\rightarrow e^{i\alpha}\chi\,,\qquad \chi^\dagger \rightarrow e^{-i\alpha}\chi^\dagger
\end{equation}
but other symmetries that are discussed below can forbid it too. 
Up to field redefinitions (and Fierz identities) there exists only one dimension-6 operator
\begin{equation}
\mathcal{O}^{(6)}= \chi^{\dagger\,2} \chi^2
\end{equation}
which results in  $\mathcal{M}\sim O(p^2)$\,.   Spinor contractions are always understood, e.g. $\chi^2=\chi^\alpha\chi_\alpha$, $\chi^{\dagger\,2}=\chi_{\dot\alpha}\chi^{\dagger\,\dot\alpha}$. 
We can make the amplitude softer than $p^2$ by forbidding the $\mathcal{O}^{(6)}$ (and the mass) by symmetries, e.g. a non-linearly realised SUSY transformation, or a fermionic shift symmetry that we discuss below, see Eq.~(\ref{eq:shiftfermion}). 
The  next dimension-8 operators which affect the soft elastic $2\rightarrow 2$ scattering must involve 4-fermions and two derivatives, resulting in the $O(p^4)$ soft behavior of the amplitude. For simplicity, we further reduce the number of operators by demanding chiral symmetry (\ref{eq:chiralsym}), so that they must contain two $\chi$'s, two $\chi^\dagger$'s, and two $\partial$'s. Up to field redefinitions there exists a unique such dimension-8 operator 
\begin{equation}
\label{eq:O8KS}
\mathcal{O}^{(8)}=-\alpha \chi^{\dagger\,2} \square\chi^2\,, 
\end{equation}
where $\square\equiv \partial^\mu \partial_\mu$ and $\alpha\in\mathbb{R}$. 
This term corresponds to the quartic term that appears in the Goldstino lagrangian within the constrained superfield formalism \cite{Casalbuoni:1988xh,Komargodski:2009rz}.  
By the field redefinition $\chi\rightarrow \chi-i\alpha (\sigma^\mu \chi^\dagger)\partial_\mu\chi^2$ and the Fierz identity $(\partial_\nu\chi^\dagger \bar{\sigma}^\rho \chi)(\partial_\mu\chi^\dagger \bar{\sigma}^\gamma \chi)=(\partial_\nu\chi^\dagger\bar{\sigma}^\rho \sigma^{\gamma}\partial_\mu\chi^\dagger)\chi^2/2$ we can map (\ref{eq:O8KS}) and the kinetic term into the standard quartic term of the Akulov-Volkov lagrangian~\cite{Volkov:1973ix} for the Goldstino, which is invariant under the non-linearly realized SUSY transformation $\chi(x)\rightarrow \chi^\prime(x^\prime)+\xi$, with $x^\prime=x+i\theta^\dagger \bar{\sigma}^\mu \xi- i\xi^\dagger \bar{\sigma}^\mu \theta$. 


The scattering amplitude for $\chi^{-} \chi^{-}  \rightarrow \chi^{-}  \chi^{-}$ and $\chi^{-} \overline{\chi}^+ \rightarrow \chi^- \overline{\chi}^+$ in the forward elastic limit reads
\begin{equation}
\mathcal{M}(s)=4\alpha s^2
\end{equation}
and hence the positivity condition (\ref{eq:positivity2}) translates into
\begin{equation}
\alpha> 0\,.
\end{equation}
Because of unitarity, crossing symmetry and analyticity, there is no non-trivial theory where $\alpha\leq 0$ in (\ref{eq:O8KS}).
For the Goldstino, $\alpha$ is set by the (inverse) SUSY breaking scale $F^2$ which must indeed be positive given the positive norm of $\sum_\alpha ||Q_\alpha|0\rangle||^2=4\langle 0| H|0\rangle=4F^2=1/\alpha$. As expected, $\alpha\rightarrow0$ corresponds to the limit $F\rightarrow \infty$ where the Goldstino decouples. 

Can we go further, i.e. forbid $\mathcal{O}^{(8)}$ and result in a softer amplitude? 
One could envision, a priori, a fermionic shift-symmetry 
\begin{equation}
\label{eq:shiftfermion}
\chi(x)\rightarrow \chi(x)+ \xi
\end{equation}
where $\xi$ is anti-commuting constant 2-spinor, in order to forbid the $\mathcal{O}^{(6,8)}$ operators and enhance the soft behavior. Such a theory would produce an EFT that starts with various dimension-10 operators, schematically of the type
\begin{equation}
\label{eq:tentativeope10}
\mathcal{O}^{(10)}\sim \partial \chi \partial \chi \partial \chi^\dagger \partial \chi^\dagger\,.
\end{equation}
The problem with this setup, and with theories that are softer than $p^4$ in general, is that a $2\rightarrow 2$ amplitude $\mathcal{M}\sim O(p^6)$ clashes with the \textit{strict positivity} of (\ref{eq:positivity2}) in any interacting theory satisfying crossing, analyticity and unitarity. 
Since the amplitude from $\mathcal{O}^{(10)}$ is of $O(p^6)$, there exists no UV completion where such a theory would be non-trivial and respect those fundamental $S$-matrix properties.
As the shift symmetry (\ref{eq:shiftfermion}) is non-linearly realised on the one-particle state generated by $\chi$, we may think of $\chi$ as a Goldstone-fermion, not dissimilar of the Goldstino.  One consequence of this no-go theorem  is that an exact fermionic symmetry (\ref{eq:shiftfermion}) can never be  restored at higher energy. Vice versa, an interacting theory in the UV which realises the fermionic symmetry linearly can not break it spontaneously and generate the associated massless Goldstone-fermion $\chi$ in the IR.\footnote{One could arrive to the same conclusion that the symmetry is never linearly realised with a different set of assumptions via the Haag-Lopuszanski-Sohnius theorem \cite{Haag:1974qh} which generalises the Coleman-Mandula theorem \cite{Coleman:1967ad}, as our fermionic symmetry (\ref{eq:shiftfermion}) is neither internal nor include spacetime translations.}

In spite of the previous argument against an exact fermionic shift symmetry (\ref{eq:shiftfermion}), it may still represent a meaningful approximate symmetry.  In principle, an \textit{arbitrarily small} explicit breaking could heal the theory and make it consistent with our fundamental requirements. It is enough that  the otherwise forbidden lower-dimensional operators, such as e.g. the mass term, turn on the subleading $O(s^2)$ coefficient in the amplitude with a positive coefficient, say $m^2s^2$.  
This is the fermionic analog of the soft-healing mechanism discussed for the longitudinal galileon modes of massive gravity in \cite{Cheung:2016yqr}.  
One can thus look at the no-go theorem from a different perspective: in order to break the fermionic symmetry in the UV, which delivers the Goldstone-fermion $\chi$ in the IR, one always needs a small explicit breaking term $\epsilon$ that makes the amplitude slightly less soft. The small parameter $\epsilon$ must be less irrelevant (or even relevant, e.g. a mass term) than the symmetric terms of the type (\ref{eq:tentativeope10}).
Should the coefficients $c_n$ or the cutoff $\Lambda$ of the higher dimensional operators $\mathcal{O}^{(n)}$ depend on the $\epsilon$ such that $c_n\rightarrow 0$ or $\Lambda\rightarrow \infty$ as we send $\epsilon\rightarrow 0$, there would be no contradiction as one would eventually recover the decoupled theory in that limit.  The most interesting case for a small but finite $\epsilon$  corresponds to  $\Lambda$ very large while the Wilson coefficients respecting the symmetry are $O(1)$, i.e. controlled by sizeable couplings $g_*=O(1)$. This case corresponds to UV completions  that enter at very large scales but are themselves non-trivial, i.e. with $g_*$ non-necessarily  small.  We will return to this supersoft behavior and possible obstructions in retaining  a finite $g_*$ in section~\ref{sec:nonunitary}.
 
 \subsection{Coupling light fields and the Goldstino}

We discuss now a neat applications of our positivity bound to the low-energy theory of a massless spin-1/2 particle coupled to other (naturally) light degrees of freedom that we may encounter in the IR: a Goldstone boson $\pi$,  a gauge boson $A_\mu$, and a massless fermion $\psi$.   
We restrict to operators that enter in the elastic $2\rightarrow 2$ scattering and follow the classification of \cite{Liu:2016idz} and \cite{Komargodski:2009rz} inspired by the theory of a Goldstino which is the prototype spin-1/2 fermion with maximally soft amplitudes. 

\subsubsection*{Coupling to fermions}

A generic theory of 2 chiral massless fermions $\chi$ and $\psi$ whose scattering amplitude is as soft as $s^2$ is fully described, up to field redefinition and Fierz identities, by two operators \cite{Komargodski:2009rz,Brignole:1997pe}:
\begin{align}
\mathcal{O}^{8}_{\chi\psi} =  -\frac{a_{\psi}}{F^2}(\chi^\dagger\psi^\dagger)\square (\chi\psi)\,,\qquad 
\widetilde{\mathcal{O}}^{8}_{\chi\psi} =&  \frac{\widetilde{a}_\psi}{F^2}(\partial_\nu\chi^\dagger\bar{\sigma}^\mu\partial^\nu\chi)(\psi^\dagger \bar{\sigma}^\mu \psi)\,,
\end{align}
where $F$ has mass dimension equal to two, $[F]=2$.
The forward elastic amplitudes for $\chi^- \psi^- \rightarrow \chi^- \psi^-$ and its crossed processes is independent of $\widetilde{a}_\psi$ and reads
\begin{equation}
\mathcal{M}(s)=\frac{a_\psi}{F^2}s^2\,,\qquad t\rightarrow 0\,.
\end{equation}
Hence, the positivity constraints (\ref{eq:positivity2}) enforced by unitarity of the underlying microscopic theory in the UV demands
\begin{equation}
a_\psi>0
\end{equation}
but leaves $\widetilde{a}_\psi$ unconstrained. For the specific case of the Goldstino,  $F$ is the SUSY breaking scale and the non-linearly realized SUSY implies $a_\psi=1$.

\subsubsection*{Coupling to Goldstone bosons}

Let consider the coupling between $\chi$ and one Goldstone boson $\pi$ which transforms as $\pi\rightarrow \pi+\mathrm{const}$:
\begin{equation}
\mathcal{O}^{(8)}_{\chi\pi}=\frac{i a_\pi}{4F^2}\partial_\mu\pi \partial^\nu \pi (\chi^\dagger \bar{\sigma}^\mu \partial_\nu\chi)+h.c.
\end{equation}
The forward elastic amplitude for the scattering $\chi^- \pi \rightarrow \chi^- \pi$ reads
\begin{equation}
\mathcal{M}(s)=\frac{a_\pi}{2F^2}s^2\,,\qquad t\rightarrow 0\,.
\end{equation}
In turn, the positivity bound (\ref{eq:positivity2}) implies 
\begin{equation}
a_\pi>0\,.
\end{equation}
A slightly weaker bound, $a_\pi\geq 0$, was derived in \cite{Dine:2009sw} by requiring luminal or subliminal propagation of the fermion excitations in a certain Goldstone boson background, along the lines  of \cite{Adams:2006sv}. We stress however that superluminality in a preferred frame (as the one set by the Goldstone boson background), as opposed to all frames, does not necessarily imply acausal propagation  or other inconsistencies like the existence of closed causal curve.  See e.g. \cite{Nicolis:2009qm} for an example with scalars where the unitarity constraint and the subluminality constraint are different, and e.g. \cite{deRham:2014lqa,Dubovsky:2007ac,Hollowood:2007kt,Papallo:2015rna} for more general discussions on the difference between acausality and superluminality.  In contrast, a violation of our positivity bound would signal unambiguously the breakdown of the rules of local and unitary quantum field theories. It is reassuring to find $a_\pi>0$ with our arguments that  have put on a firm ground this inequality and, in turn, the bound on the superpotential of \cite{Dine:2009sw}.

\subsubsection*{Coupling to gauge bosons}

We finally consider the coupling between $\chi$ and a $U(1)$ gauge boson $A_\mu$. As we are after $O(s^2)$ terms, we require $\chi$ to be neutral under the $U(1)$ so that $A_\mu$ and $\chi$ have only dipole or multipole interactions via the the field strength $F_{\mu\nu}=\partial_\mu A_\nu -\partial_\nu A_\mu$: 
\begin{equation}
\mathcal{O}^{(8)}_{\psi A}=-\frac{i a_{A}}{2F^2}(\chi^\dagger \bar{\sigma}^\mu \partial_\nu\chi) F_{\mu\rho}F^{\nu\rho}+h.c.
\end{equation}
This  operator gives rise to the following forward elastic scattering amplitude
\begin{equation}
\mathcal{M}(s)=\frac{a_{A}}{F^2} s^2
\end{equation}
for the process $\chi^- A^{\pm}\rightarrow \chi^- A^{\pm}$, and hence
\begin{equation}
a_{A}>0
\end{equation}
in any theory where $\chi$ and $A$ interact with each other. 

\section{Supersoft amplitudes}
\label{sec:nonunitary}

We now look for unitary theories that are even softer than $O(p^4)$ and yet respect the positivity (\ref{eq:positivity2}).
As we have seen in the previous section, such a supersoft behavior can not be enforced in a strict sense in any interacting theory as the coefficient of $s^2$ in the amplitude would need to be strictly positive. 
However, since this constraint could be satisfied for an \textit{arbitrarily small} (and positive) coefficient, we may ask whether supersoft theories can make sense as a limiting case of unitary theories where the $O(p^4)$-terms in the amplitude are taken smaller and smaller, say by a symmetry.  
As we will see, for certain class of theories there exists an obstruction in retaining a non-trivial interacting UV completion at the cutoff when the $O(p^4)$-terms are made very small.
In order to illustrate these points we first study the case of a supersoft spin-1/2 fermion. 

\subsection{Fermionic shift symmetry}
\label{subsec:fermshiftsym}

We consider a chiral spin-1/2 fermion field endowed with a shift symmetry 
\begin{equation}
\label{eq:shiftfermionbis}
\chi(x)\rightarrow \chi(x)+ \xi
\end{equation}
that forbids $O(p^4)$ term in the $2\rightarrow 2$ scattering. 
We assume a one-scale ($\Lambda$) one-coupling ($g_*$) power-counting scheme  
\begin{align}
\label{eq:1g1S}
\mathcal{L} = \chi^\dagger i\bar{\sigma}^\mu\partial_\mu\chi+ \frac{\Lambda^4}{g_*^2}L[\frac{g_*\partial^{n}\chi}{\Lambda^{3/2+n}},\epsilon\frac{g_*\chi}{\Lambda^{3/2}} ] 
\end{align}
where $L$ is a dimensionless function and $n\geq1$.   Terms that break the fermionic shift symmetry  cost the insertion of a small spurion $\epsilon\ll 1$ ~\footnote{This power counting can be realized, e.g.,  by   having $\partial_\mu\chi$  linearly coupled to a spin-3/2 operators $\mathcal{O}^\mu_\alpha$ of a strong sector, i.e. $\lambda \partial_\mu \psi^\alpha \mathcal{O}^\mu_\alpha$, which is eventually integrated out after it develops a mass gap, analogously to the partial compositeness scenario.  The terms with $\epsilon$ are generated instead by small corrections that break the shift symmetry such as e.g.  $\epsilon \chi \mathcal{O}$, where $\mathcal{O}$ is a spin-1/2 operator of the strong sector like in ordinary partial compositeness to generate $\chi$'s mass. See e.g. \cite{Bellazzini:2014yua,Panico:2015jxa} for a review on partial compositeness.}
\begin{equation}
\label{pcountingrem}
\mathcal{L}=\chi^\dagger i\bar{\sigma}^\mu\partial_\mu\chi +\epsilon^4 c_1 \frac{g_*^2}{\Lambda^2}\chi^{\dagger\,2}\chi^2- \epsilon^2 c_2 \frac{g_*^2}{\Lambda^4} \chi^{\dagger\,2} \square\chi^2 + c_3 \frac{g_*^2}{\Lambda^{6}}(\partial_\nu \chi^\dagger \partial^\nu \chi^\dagger)(\partial_\mu \chi \partial^\mu \chi) +\ldots 
\end{equation}
whereas the symmetric terms are not suppressed.
We are not including the mass for simplicity, as it can be forbidden e.g. by chiral symmetry (\ref{eq:chiralsym}) which has its own separate spurion $\epsilon_m$, set here to zero.  Aside the shift symmetry, for $g_*=4\pi$ one has the traditional power counting, or na\"ive dimensional analysis (NDA) \cite{Manohar:1983md}, of a full-fledged strongly coupled theory at the scale $\Lambda$. In such a case, the only available expansion parameter is $E/\Lambda$ and thus $c_i=O(1)$. Smaller $g_*$ extends the NDA  since the theory admits also a perturbative expansions in $g_*^2/(16\pi^2)$ which counts the number of  loops relative to the leading classical contribution to the observables \cite{Giudice:2007fh}. Under this latter assumption, it is meaningful to classify the operators based on their size that can be either tree- or loop-level $c_i=O((g_*^2/16\pi^2)^\ell)$, corresponding to $\ell=0$ and $\ell\geq1$ respectively. Note that $g_*$ corresponds to the coupling at the scale $m_*$ and below; it does not represent the original microscopic coupling at higher energy should the theory emerge  in the IR from a strong sector. 
The scale $\Lambda$, which controls the derivatives expansion, usually corresponds to an actual physical threshold, e.g. the mass of a new particle exchanged in the $\chi\chi$-scattering. If this is so,  the EFT breaks down at $E\sim \Lambda$, independently on the size of $g_*$, since the new on-shell degree of freedom must be added to the spectrum. The coupling $g_*$ tells us whether this new entry is strongly or weakly coupled.
We come back later to the exceptional case where $\Lambda$ does not correspond to a physical threshold, meaning that $\mathcal{L}$ is actually a finite polynomial in derivatives and fields.
 
 The $2\rightarrow 2$ scattering amplitude scales as 
\begin{equation}
\label{ampFG}
\mathcal{M}\sim g_*^2\left(\frac{E}{\Lambda}\right)^{6}\left[\left(c_3+\ldots \right)+ c_2\left(\frac{\epsilon\Lambda}{E}\right)^{2}+ c_1\left(\frac{\epsilon\Lambda}{E}\right)^{4}+ o\left(\frac{E}{\Lambda}\right)^{2}\right]
\end{equation}
where $E$ is the typical energy at hand. The ellipses $\ldots$ refer to other symmetric $O(p^6)$-terms that we have omitted for simplicity; hereafter $c_3$ is a short-hand for all such contributions. 
We take $c_2>0$ to comply with the positivity bound for any finite $\epsilon$.
 The supersoft term $c_3$ in (\ref{pcountingrem}) dominates over the other terms whenever
 \begin{equation}
\Lambda_{IR}\equiv \left(\frac{c_2}{c_3}\right)^{1/2} \Lambda \epsilon \ll E \ll \Lambda\,.
 \end{equation}
   It fails to dominate only in the tiny window $0<E< \Lambda_{IR}$ that we could apparenly shrink to zero arbitrarily as $\epsilon\rightarrow 0$.
  For a fixed coupling $g_*$ one may even continuously reach $\epsilon=0$ by making $\Lambda$ dependent on $\epsilon$, $\Lambda=\Lambda(\epsilon)$, and requiring that it grows arbitrarily slowly for $\epsilon\rightarrow 0$, in order to recover the decoupled theory when the amplitude becomes strictly softer than $s^2$. Taking e.g. $\Lambda\sim \log\epsilon$ or $\Lambda\sim 1/\epsilon^{n}$ with $0<n\ll 1$, we can treat the symmetry breaking term as a small perturbation effectively to all energies 
\begin{equation}
\label{eq:Lambdaepsi}
\Lambda\rightarrow\infty \qquad \mbox{while }\qquad \Lambda_{IR}\rightarrow 0.
\end{equation}
For finite $\epsilon$ and $\Lambda$, with  $\Lambda\epsilon \ll \Lambda$, the symmetry-preserving terms are important in most of the range of validity of  the EFT.  
This setup realizes an even more extreme version of the ``remedios'' power counting proposed in \cite{Liu:2016idz}, as the leading amplitudes of $O(p^6)$ is generated by dimension-10 operators that dominate the lower dimensional ones at low-energy, while retaining a sensible EFT scheme.  

The fact that a small breaking $\epsilon$ can affect the cutoff $\Lambda$ and thus heal the theory was loosely inspired by the case of the dilaton where the scale of conformal symmetry breaking $f$ can be stabilised  moving away from exact conformality, e.g. with an almost marginal perturbation which delivers a decoupled dilaton, $f\rightarrow \infty$, when $\epsilon\rightarrow0$. For the dilaton, however, the decoupling of $f$ is exponentially fast, while here we demand logarithmic sensitivity. 
More generally, one can imagine a barely stable/unstable configuration in the UV that is stabilised in a healthy theory by a small perturbation which imprints itself in the non-analytic dependence upon $\epsilon$ of the cutoff, and possibly of the amplitude. 
 \subsection{Obstructions, loopholes and massive gravity}
 \label{obstLoop}

The picture described in section~\ref{subsec:fermshiftsym} seems to allow amplitudes that are practically, but not exactly, as soft as $O(p^6)$ while respecting the positivity bound (\ref{eq:positivity2}).
In this subsection we want to point out that it may actually exist an obstruction in taking $\epsilon$ arbitrarily small while retaining a finite  coupling  $g_*$ at the scale $\Lambda$. 

The dispersion relation (\ref{eq:dispn}),  that in the case at hand reads
\begin{equation}
\label{eq:disp2}
\mathcal{M}^{\prime\prime}(0) =  \frac{2}{\pi}\int^\infty_{0} \frac{ds}{s^2}\left[ \sigma^{\mathrm{tot}}(\chi  \chi \rightarrow \mathrm{anything})(s)+\sigma^{\mathrm{tot}}(\bar{\chi} \chi\rightarrow \mathrm{anything})(s)\right]\,,
\end{equation} 
  can be used to set an upper limit on the value of the cutoff \cite{Nicolis:2009qm}. Indeed, one can use the dispersion relation to determine the scale $\Lambda_*$ where the low-energy discontinuity on the right-hand side of (\ref{eq:disp2}) can no longer match the value on left-hand side, that is the ultimate scale where a new non-analyticity---i.e. the threshold of new degrees of freedom---is required to kick-in.  
Assuming $\epsilon^2\ll 1$, the consistency between the two sides of (\ref{eq:disp2}) demands\footnote{We are not showing the dependence on the  $c_i$ just for easy of presentation. Including Wilson coefficients $c_i$ results into multiplying the right-hand side of Eq.~(\ref{eq:boundeps}) by the factor $(c_3^2/c_2)$.} that $\epsilon^2 g_*^2/\Lambda^4 \sim  (g_*^4/16\pi^2) \Lambda_{*}^8/\Lambda^{12}$  and hence 
\begin{equation}
\label{boundesp}
 \Lambda_* \sim \Lambda \times \left(\frac{4\pi\epsilon}{g_*}\right)^{1/4}\,. 
\end{equation}
For $\epsilon\ll1$, the $\Lambda_*$ becomes much smaller than $\Lambda$ (and yet bigger than $\Lambda_{IR}$) which is consistent with the way we performed this calculation. However this is not consistent with our one-scale power counting  where  $\Lambda$ was supposed to be  controlling the derivatives expansion in (\ref{pcountingrem}), setting its radius of convergence i.e. the threshold for the new states. This power-counting does not tolerate hierarchically separated $\Lambda$ and $\Lambda_*$ and this puts a lower bound on $\epsilon$. Requiring that $\Lambda_*\sim  \Lambda$ and using  (\ref{boundesp}) one extracts the estimate  
\begin{equation}
\label{eq:boundeps}
\epsilon^2 \sim \left(\frac{g_*^2}{16\pi^2}\right)\,. 
\end{equation}
That is, $\epsilon^2$ can not be much smaller than a one-loop factor (relative to $c_3^2/c_2$ should we restore the Wilson's coefficients dependence),  
and as $\epsilon\rightarrow0$ so does $g_*$.  The UV completion would thus appear weakly coupled, i.e. a perturbation of the free theory which is infinitely soft and yet perfectly healthy. 

This argument is quite generic but the resulting constraint (\ref{eq:boundeps}) should be regarded only as a na\"ive estimate:  as $\Lambda_*$ approaches  $\Lambda$ from below, higher derivatives terms become gradually more important for evaluating the low-energy contribution to the right-hand side of the dispersion relation (\ref{eq:disp2}). One should thus replace the fractional powers in (\ref{boundesp}) with even smaller ones, resulting in a poor sensitivity on the smallness  of $\epsilon$. 
For $\Lambda_*$ as large as $\Lambda$ or bigger we can not even make a reliable calculation nor present an argument for a lower bound on $\epsilon$. 

There exist nonetheless an exceptional class of theories where the calculation can be done in principle for $\Lambda_*\ll \Lambda$ (and even for $\Lambda_* > \Lambda$ but still below the strong coupling scale $\Lambda_{strong}=\Lambda\times (4\pi/g_*)^{1/3}$).  
It is e.g. conceivable a derivatives series (\ref{pcountingrem}) whose leading order operators terminate in a finite order polynomial, such that we do not need to identify the parameter $\Lambda$ with an actual physical threshold associated with non-analytic behaviour.  The Galileon \cite{Nicolis:2008in} is one such a theory because it enjoys a non-renormalization theorem for a finite set of operators $\mathcal{L}_{2,\ldots, 5}$. Higher derivatives are eventually generated but they are suppressed relative to the non-renormalized terms by loop factors $(g_*^2/16\pi^2)^\ell$, for moderate coupling $g_*$.
In this class of theories, we can take $\Lambda_{IR}\ll \Lambda_*\ll \Lambda$ without running into an apparent inconsistency with the dispersion relation above.

These theories would look essentially supersoft in the window $\Lambda_{IR} \ll E \ll \Lambda_* $ should the lower end go to zero much faster than its upper end, as $\epsilon\rightarrow 0$.
But in fact, this can not  actually happen. Indeed, $g_*$ and $\Lambda$ are no longer physical quantities in these exceptional theories when $\epsilon\ll1$, since it makes no sense to extrapolate  the EFT amplitudes at energies above $\Lambda_*$.  
A more physical definition of coupling constant controlling the UV completion is for example the value of the $2\rightarrow 2$ amplitude at $\Lambda_*$, 
\begin{equation}
\label{ampFGbis}
\mathcal{M}= \widetilde{g}_*^2\left(\frac{E}{\Lambda_*}\right)^{6}\left[c_3 +c_2\left(\frac{\widetilde{g}_*^2}{16\pi^2}\right)\left(\frac{\Lambda_*}{E}\right)^{2}+ c_1\left(\frac{\widetilde{g}_*^2}{16\pi^2}\right)^2\left(\frac{\Lambda_*}{E}\right)^{4} \right]
\end{equation}
where  $\Lambda/\Lambda_*=\widetilde{g}_*/(4\pi\epsilon)$ and we have defined $\widetilde{g}^2_*= g_*^{2}(4\pi\epsilon/g_*)^{3/2}$. 
One can see that the ordinary soft terms $c_{1,2}$ appear again  with  only a $1$-loop suppression factor compared to the supersoft $c_3$, when we express the amplitude in terms of the physical coupling $\widetilde{g}_*$.  Equivalently, the ratio $\Lambda_{IR}/\Lambda_* = \widetilde{g}_*/4\pi$ is not arbitrarily large when $\epsilon\rightarrow 0$.
 Again, this argument holds only for $\Lambda_*< \Lambda_{strong}$, i.e. for UV completions that enter before the onset of  the fully strong coupling regime where calculability is completely lost. It does apply though  for strongish, $O(1)$, couplings. 

The Galileon \cite{Nicolis:2008in} is an example of supersoft theory defined by a finite set of operators that admit various perturbations, e.g. the conformal Galileon deformation \cite{Nicolis:2009qm,Nicolis:2008in} or the coupling to gravity \cite{Pirtskhalava:2015nla,Deffayet:2009wt}, that turn on the $O(p^4)$-terms. 
In the ghost-free massive gravity (see e.g. \cite{deRham:2010ik,deRham:2010kj,Hinterbichler:2011tt} and references therein), the Galileon describes the scalar polarization that gives the $O(p^6)$-behavior to the scattering amplitudes of massive gravitons.  The finite graviton mass $m_g$ generates  a subleading $O(p^4)$-term \cite{Cheung:2016yqr}.
More specifically, the leading Galileon amplitude scales as $g_*^2 (E/\Lambda)^6$ where $\Lambda^6=g_*^2 m_g^4 M^2_{Planck}$, while 
a finite graviton soft mass generates a subleading $g_*^2 m_g^2 E^4/\Lambda^6$ contribution.  Making thus the identifications $\epsilon^2=m_g^2/\Lambda^2$, the ratio 
 \begin{equation}
 \label{eq:boundMassiveG}
 \left(\frac{m_g}{  g_* M_{Planck}}\right)^2\sim   \left(\frac{g_*^2}{16\pi^2}\right)^3
 \end{equation}
 is expected to be around three loops, given the general obstruction on the size of $\epsilon$ within our  power counting. This ratio can still be taken very small but only in the trivial way $g_*\ll 1$, too.~\footnote{The traditional decoupling limit in massive gravity corresponds to $M_{Planck}\rightarrow\infty$ holding $\Lambda_3=\Lambda/g_{*}^{1/3}=(m_g^2 M_{Planck})^{1/3}$ fixed. The relation (\ref{eq:boundMassiveG}) requires that $g_*$ is actually vanishing too, scaling as $g_*\sim(4\pi m_g/\Lambda_3)^{3/4}$. The $O(p^6)$-term survives in this limit as it scales only with $\Lambda^{-6}_3$.}  
 This relation implies also that $\Lambda$ is at most one-loop suppressed relative to $g_* M_{Planck}$.

\section{Conclusions}\label{sec:conc}

In this paper we have studied how crossing symmetry, unitarity and analyticity of  a microscopic theory translate into positivity bounds of the resulting EFT in the IR. 
We have proved that those fundamental requirements imply the strict positivity of the leading $O(s^2)$-terms in the elastic forward scattering amplitudes for particles with arbitrary spin at low-energy.
In turn, we have shown that EFT's  that produce amplitudes strictly softer than $O(p^4)$ do not admit UV completions that satisfy the basic set of assumptions of a scattering theory.
These results are based on the analytic continuation of the pure states' density matrices, i.e. the spin projectors, that are traced with the amputated correlators to provide the elastic forward amplitudes.

For highly boosted particles with energy below the cutoff, one can reinterpret these soft bounds as restrictions on the rate of growth in energy of scattering amplitudes, within the validity of the EFT. 
We have studied in detail the example of a chiral spin-1/2 fermion that saturates this limit, reproducing essentially the coupling structure of the Goldstino from SUSY breaking. While the positivity constraints are trivially satisfied by the self-interactions dictated by the Akulov-Volkov effective action, they impose non-trivial conditions on the couplings to the other light degrees of freedom.

We have also shown how to make sense of theories with amplitudes that are loosely, as opposed to strictly, softer than $O(p^4)$. They should be understood as the limiting case of unitary theories where the operators that control  $O(p^4)$-terms in the amplitudes are (possibly arbitrarily) suppressed by a symmetry, while the cutoff that controls even softer corrections is taken (arbitrarily) large, but at a much slower rate. Since the supersoft terms  decouple much more slowly than the $O(p^4)$-terms, one effectively obtains a supersoft theory to almost all energies below the cutoff.
For exceptional EFTs that contain only a finite set of derivatives, such as e.g. the Galileon,  we have argued there exists an obstruction in having non-trivial, i.e. non-weakly-coupled, UV completions at the scale $\Lambda$. 
 
We have discussed in detail the supersoft theory of a chiral spin-1/2 fermion with a fermionic shift symmetry $\chi\rightarrow \chi+\xi$ that would forbid $O(p^4)$-terms in the amplitudes, but which is perturbed by naturally small Goldstino-like interactions, as well as by the couplings to other massless particles. The positivity constraints can be satisfied requiring that the cutoff itself grows slowly as the spurion associated to the breaking of the shift symmetry is taken to arbitrary small values.  
 
\section*{Acknowledgments}

I thank Riccardo Torre for the involvement and support at the early stage of this project. I thank Riccardo Torre, Grant Remmen and Cliff Cheung  for reading and commenting on the paper.  I am grateful to Francesco Riva, Enrico Trincherini and Riccardo Rattazzi for the stimulating discussions. I thank the LPTHE, Marco Cirelli and the ERC Starting Grant 278234 “NewDark” for the hospitality and support while this work was in progress.
This work is supported in part by the MIUR-FIRB grant RBFR12H1MW. 

\appendix

\section{Polarizations}
\label{app:polarizations}

The polarizations $u_\ell$ and $v_\ell$ are defined by the overlaps (\ref{overlap}) and (\ref{overlapLoc}) between a one particle state $|p^\sigma_a \rangle$ and $\Psi^\dagger_{\ell\, a}(x)| 0\rangle$, or between the anti-particle $|p^\sigma_{\bar{a}} \rangle$ and $\Psi_{\ell\, a}(x)| 0\rangle$. We quickly review below their basic properties following \cite{Weinberg:1995mt}. As they are slightly different for massive and massless particles, we discuss them in turn.

\subsubsection*{Massive fields}
Under a Lorentz transformation $\Lambda$, a one particle massive state transforms with an irreducible unitary representation $\mathcal{L}$ of the little group $SU(2)\sim SO(3)$, see Eq.~\ref{eq:lorentzT}, whereas the field $\Psi_\ell$ transforms covariantly, i.e. $U(\Lambda)\Psi(x)U^\dagger(\Lambda)=D(\Lambda^{-1})_{\ell\ell^\prime }\Psi_{\ell^\prime}(\Lambda x)$, according to some (generically non-unitary) representation $D\in (A,B)$ of $SU(2)_A\times SU(2)_B\sim SO(3,1)$ where the index $\ell=(\alpha\beta)$ collectively labels the states in the representation. 
The Wigner rotation $W(\Lambda,p)=L(\Lambda p)^{-1}\Lambda L(p)\in SO(3)$ is defined in terms of the standard Lorentz transformation $L(p)$ that sends the little-group reference vector $k_r=(m,0,0,0)^T$ to $p$. 
From the overlaps (\ref{overlap}) and (\ref{overlapLoc}), wee see that the polarizations are charged under $\mathrm{Lorentz}\times (\mathrm{Little-Group})$,
\begin{align}
\label{wfdef1}
D_{\ell \ell^\prime}(\Lambda) u^\sigma_{\ell^\prime}(\p) =  u^{\sigma^\prime}_\ell (\Lambda p) \mathcal{L}_{\sigma^\prime \sigma}(W(\Lambda,p))\,,\qquad D_{\ell \ell^\prime}(\Lambda) v^\sigma_{\ell^\prime}(\p) =  v^{\sigma^\prime}_\ell (\Lambda p) \mathcal{L}^*_{\sigma^\prime \sigma}(W(\Lambda,p))\,.
\end{align}
Lorentz is acting on the right-hand side while the little-group on the left-hand side. 
Clearly, for a real representation $D\sim D^*$ one can choose $v_\ell\sim u^*_\ell$. For example, in the vectorial spin-1 representation $(1/2,1/2)$ it is customary to choose $\epsilon_\mu=u_\mu=v^*_\mu$, the so-called helicity basis.  Analogously, the 4-component spinors $(1/2,0)\oplus(0,1/2)$ with $u_\alpha=v^*_\alpha$ define the Majorana basis.

Locality implies that under a $CPT$ transformation, which is realized by an anti-unitary operator $U_\mathrm{CPT}$, we can always choose the overall phases such that 
\begin{equation}
U_{\mathrm{CPT}} |p^\sigma_a\rangle= (-1)^{S+\sigma} |p^{-\sigma}_{\overline{a}}\rangle\,, \qquad U_{\mathrm{CPT}}\Psi_\ell(x) U^{-1}_{\mathrm{CPT}}=(-1)^{2B} \Psi^{\dagger}_\ell(-x)\,.
\end{equation}
Therefore, particles and anti-particles polarizations defined by the overlaps (\ref{overlap}) and (\ref{overlapLoc}) must be related by 
\begin{equation}
\label{eq:generalCPTwave}
u^\sigma_{\ell}(\p)=(-1)^{2B+S+\sigma}v_{\ell}^{-\sigma}(\p)\,.
\end{equation}
 As $\sigma$ runs from $-S$ to $S$,  the phase is always real i.e. either $\pm1$.
For the vector and Dirac representation one recovers the relations given in  Eq.~(\ref{eq:uvgamma5}) that one can explicitly check against Eq.~(\ref{expliciteps}), Eq.~(\ref{explicitspin}) and the literature, e.g. \cite{Dreiner:2008tw}. This implies  that the massive density matrices  of particles and anti-particles are related:
\begin{equation}
\label{Eq:massiverhosrels}
u^\sigma(\p) u^{\sigma\,\dagger }(\p)=v^{-\sigma}(\p) v^{-\sigma\,\dagger }(\p)\,,
\end{equation}
that is $\rho^\sigma= \widetilde{\rho}^{-\sigma}$.
The Dirac representation $(1/2,0)\oplus (0,1/2)$ is reducible and the $(-1)^{2B}$ factor in (\ref{eq:generalCPTwave}) gives rise to the  $\gamma_5$ of Eq.~(\ref{eq:uvgamma5}), which in turn gives  
\begin{equation}
u^\sigma(\p) u^{\sigma\,\dagger }(\p)=\gamma^5 v^{-\sigma}(\p) v^{-\sigma\,\dagger }(\p)\gamma^5
\end{equation}
 for massive 4-component fermions, in agreement with (\ref{densfermion2}). In terms of $\bar{u}$ and $\bar{v}$ it reads $u^\sigma(\p) \bar{u}^{\sigma}(\p)=-\gamma^5 v^{-\sigma}(\p) \bar{v}^{-\sigma}(\p)\gamma^5$ as $\{\gamma^0,\gamma^5\}=0$.
\\

Eq.(\ref{wfdef1}) provides also  a constructive definition for the polarizations: by setting $\p=0$ and $\Lambda$ equal to the Lorentz transformation $L(p)$ we have $W=1$, and the polarizations  are obtained by the ones in the rest frame of the particle with $\p=0$, namely
\begin{equation}
\label{defuL}
u^{\sigma}_\ell (\p)= D_{\ell \ell^\prime}(L(p)) u^\sigma_{\ell^\prime}(\mathbf{0}) \,,\qquad v^{\sigma}_\ell (\p)= D_{\ell \ell^\prime}(L(p)) v^\sigma_{\ell^\prime}(\mathbf{0})\,.
\end{equation}
Moreover, under an arbitrary rotation $\Lambda=\mathcal{R}$ we have $W=\mathcal{R}$ for any $p$ and
\begin{equation}
\label{defuR}
D_{\ell \ell^\prime}(\mathcal{R}) u^\sigma_{\ell^\prime}(\mathbf{0}) = u^{\sigma^\prime}_\ell (\mathbf{0}) \mathcal{L}_{\sigma^\prime \sigma}(\mathcal{R})\,.
\end{equation} 
Since we are defining one particle states at rest with the definite spin along the $z$-axis, that is $\mathcal{L}_{\sigma^\prime \sigma}(\mathcal{R}(\hat{\mathbf{z}}))=\mathrm{Exp}[i\sigma \theta] \delta_{\sigma^\prime \sigma}$, 
the $u^\sigma_{\ell^\prime}(\mathbf{0})$ can be look for studying the eigenvectors of $z-$rotations generated by $D(\mathbf{J}^z)$ 
\begin{equation}
D(\mathbf{J}^z)_{\ell\ell^\prime } \, u^\sigma_{\ell^\prime}(\mathbf{0})=\sigma u^\sigma_{\ell}(\mathbf{0})\,.
\end{equation}
For example, a massive spin-1 state created by the vector $(1/2,1/2)$ has $D(\mathbf{J}^z)_{ij}=-i \epsilon_{3ij}$ (and $D(\mathbf{J}^z)_{00}=D(\mathbf{J}^z)_{i0}=D(\mathbf{J}^z)_{i0}=0$); hence the following vectors are a valid choice of  polarizations 
\begin{equation}
\label{expliciteps}
\mbox{vector:} \qquad  \epsilon_\mu^{\pm}(p)= \mp R(\hat{\p})
\left(
\begin{array}{c}
0\\
 1/\sqrt{2}  \\
\pm i/\sqrt{2} \\
0
\end{array}
\right)\,,\qquad \epsilon_\mu^{0}=
R(\hat{\p})
\left(
\begin{array}{c}
 p^z/m  \\
0\\
0\\
E/m
\end{array}\right)  \,,
\end{equation}
where $L(p)=R(\hat{\p})B(|\p|)$ with $B(|\p|)$ a boost on the $z-$axis (which has no effect on the $\epsilon_\mu^{\pm}(\mathbf{0})$).  $R(\hat{\p})$ is a rotation that aligns the $z$-axis along $\p$.
Note that 
\begin{equation}
\label{eq:polspin1}
p^\mu \epsilon_\mu^{\sigma}(\p)=0\,, \qquad\epsilon^{\sigma\,*}_\mu(\p) \epsilon^{\sigma^\prime}_\mu(\p)=-\delta^{\sigma\sigma^\prime}\,,\qquad \epsilon^{\sigma\,*}_\mu(\p)=(-1)^\sigma \epsilon^{-\sigma}_\mu(\p)
\end{equation}
 In fact, one could have used these relations as definition of vector polarizations.  
 
 Other useful contractions in the $2\rightarrow 2 $ scattering of identical spin-1 particles of momenta $p_{1,2}$ in the c.o.m frame and moving along the $z$-axis are  
 \begin{equation}
 \label{eqspin1contrvar}
 p^\mu_i \epsilon^{\pm}_{\mu}(\p_j)=  0\,,\qquad \epsilon^{0\,\mu}(\p_1) \epsilon^{0}_{\mu}(\p_2) =  -\frac{(s-2m^2)}{2m^2}\,,\qquad \epsilon^{0}_\mu(\p_1) p_{2}^\mu =\frac{\sqrt{-u s}}{2m}\,,\qquad \epsilon^{0\,\mu}(\p_1) \epsilon^{\pm}_\mu(\p_2)=0\,.
 \end{equation}

One could repeat the same derivation for a massive spin-1/2 in the $(1/2,0)\oplus (0,1/2)$ but it is faster to start from the Dirac equations, $(\slashed{p}-m)u^\sigma(\p)=(\slashed{p}+m)v^\sigma(\p)=0$, that are easily solved in the rest frame, i.e.  $u^\sigma(\mathbf{0})= \sqrt{m}( \xi^\sigma, \xi^\sigma)^T$ and $v^{\pm}(\mathbf{0})=\sqrt{m}( \pm \xi^{\mp}, \mp \xi^{\mp})^T$ ,  and then boosted  to 
\begin{equation}
\label{explicitspin}
\mbox{Dirac spinor:} \qquad  u^\sigma(\p)= 
\left(
\begin{array}{c}
 \sqrt{p \cdot \sigma} \xi^\sigma  \\
 \sqrt{p \cdot \bar{\sigma}} \xi^\sigma
\end{array}
\right)\,,
\qquad 
v^\sigma(\p)=\left(
\begin{array}{c}
 \sqrt{p \cdot \sigma} \eta^\sigma  \\
 - \sqrt{p \cdot \bar{\sigma}} \eta^\sigma
\end{array}
\right)\,,
\end{equation}
where $\xi^\sigma$ and $\eta^\sigma$ are a 2-component spinor, $\xi^\sigma=(1,0)^T$,  $\xi^-=(0,1)^T$ and $\eta^\sigma=-i\sigma^2\xi^{\sigma\,*}$.
All the independent traces that one takes with the density matrix $\rho^\sigma(\p)=u^\sigma(\p) u^{\sigma\,\dagger}(\p)$ (as well as with $\widetilde{\rho}$ for the anti-particles) can be recast  in terms of the 4-momentum $k_\mu$ and the polarization 4-vector $a_{\mu}(k)$ that appears in Eq.~(\ref{densfermion1}). Indeed,  each trace in $\Tr\left[\rho^\sigma(\p) \gamma^0\cdot \{1,\gamma^5, \gamma^\mu, \gamma^\mu \gamma^5, [\gamma^\mu,\gamma^\nu]\}\right]$ corresponds to one of the following bilinears 
\begin{align}
\label{tracesgammaA}
\bar{u}^\sigma(\k) \gamma^\mu u^\sigma(\k)=2k^\mu\,, \qquad  \bar{u}^\sigma(\k)\gamma_\mu\gamma^5 u^\sigma(\k)=2 m a^{\sigma}_\mu(\k)\,, \\
\bar{u}^\sigma(\k) u^\sigma(\k)=2 m\,,\qquad \bar{u}^\sigma(\k) \gamma^5 u^\sigma(\k)=0\,,\qquad  \bar{u}^\sigma(\k) [\gamma^\mu,\gamma^\nu] u^\sigma(\k)=4 i\epsilon^{\mu\nu\alpha\beta}k_\alpha a^{\sigma}_\beta(\k)\,.
\end{align} 
One can actually use the right-most expression in Eq.~(\ref{tracesgammaA}) as definition of $a^\sigma_\mu$, which thus implies the general parametrization~(\ref{densfermion1}) for the density matrix.  In the rest frame, the polarization 4-vector of pure states reduces to $a^{\pm}_\mu=(0,0,0,\pm 1)_\mu$ as one can check directly using $u(\mathbf{0})$ given above. Boosted along the $z-$direction by velocity $\beta=-k^z/k^0$ it becomes $ a_\mu^{\pm}(\k)=\pm\frac{1}{m_\Psi}(k^z, 0,0,k^0)^T$. In any frame $a_\mu a^\mu=-1$ (for pure states) and $a_\mu k^\mu=0$. 

Since the rotations are the diagonal subgroup $SU(2)_{A+B}$ of $SU(2)_A\times SU_B(2)$, they are generated by $\mathbf{J}=\mathbf{J_A} + \mathbf{J}_B$, meaning that  Eq.~(\ref{defuR}) implies 
\begin{equation}
D(\mathbf{J}_A)_{\alpha\alpha^\prime} u^\sigma_{\alpha^\prime \beta}(\mathbf{0}) + D(\mathbf{J}_B)_{\beta\beta^\prime} u^\sigma_{\alpha \beta^\prime}(\mathbf{0})=\mathcal{L}(\mathbf{J}_S)_{\sigma^\prime\sigma} u^{\sigma^\prime}_{\alpha \beta}(\mathbf{0})
\end{equation}
and analogous for $v^\sigma_{\alpha\beta}$ up to sending $\mathcal{L}(\mathbf{J}_S)_{\sigma^\prime\sigma}\rightarrow -\mathcal{L}^*(\mathbf{J}_S)_{\sigma^\prime\sigma}$ .  In other words, $u^\sigma_{\alpha\beta}$ is proportional to the $SU(2)$ Clebsch-Gordan coefficient for the spin $J=S$ inside $\mathbf{r}_A\otimes \mathbf{r}_B=\bigoplus_{J=|A-B|}^{J=A+B} \mathbf{r}_J$ 
\begin{equation}
u_{\alpha\beta}^\sigma(\mathbf{0}) \propto C_{(AB)\alpha\beta}^{(S)\sigma}\,,
\end{equation}
and analogous for $v^\sigma_{\alpha\beta}$. The Clebsch-Gordan coefficients are unitary, i.e. $\sum_\sigma C_{(AB)\alpha\beta}^{(S)\sigma} C_{(AB)\gamma\delta}^{(S)\sigma\,*}=\delta_{\alpha\gamma}\delta_{\beta\delta}$ and $\sum_{\alpha\beta} C_{(AB)\alpha\beta}^{(S^\prime)\sigma^\prime\,*} C_{(AB)\alpha\beta}^{(S)\sigma} =\delta_{S S^\prime}\delta^{\sigma\sigma^\prime}$. Thus, the sum over $\sigma$  of the density matrix in the particle's rest frame $\sum_\sigma u_{\alpha\beta}^\sigma(\mathbf{0})u_{\gamma\delta}^{\sigma\,*}(\mathbf{0})$ is nothing but the projector on the spin $S$ in the basis labeled by the pair of indexes $\alpha$ and $\beta$ (up to an overall constant). For example, in the vector representation $\Pi_{\mu\nu}(\mathbf{0})=\sum_\sigma \rho^\sigma_{\mu\nu}(\mathbf{0})$ is proportional to projector $\eta_{\mu\nu}-k_\mu k_\nu/M^2$ orthogonal to $k=(M,0,0,0)^T$ in the 4-vector space where the time direction is a 3D scalar, while $\Pi_{\mu\nu}(\p)=\sum_\sigma \rho^\sigma_{\mu\nu}(\mathbf{p})$ is proportional to the projector $\eta_{\mu\nu}-p_\mu p_\nu/M^2$ orthogonal to $p=L(p)k$.

The sum over $\sigma$ of the density matrices, which is nothing but the numerator of the propagator of $\Psi_\ell$ in Fourier space, can be written for irreducible representations as a polynomial $P(p)$ of the whole 4-momentum $p$ of order $2S$, with $P(-p)=(-1)^{2S}P(p)$ \cite{Weinberg:1995mt,Weinberg:1964cn,Weinberg:1969di}. This relation ensures causality, i.e. vanishing (anti-)commutator between spacelike asymptotic free fields with (half-)integer spins. For the Dirac representations, which is reducible, it becomes $P(-p)=-\tilde{P}(p)$ where $\tilde{P}$ is the sum over the polarizations of the anti-particle. In section~\ref{sec:frowardAndCross} we proved that this property actually hold spin by spin, i.e. without even summing over the $\sigma$.

\subsubsection*{Massless fields}

Massless fields need somewhat more care as one has to extend the notion of  ordinary Lorentz representations to the case of gauge fields which realise Lorentz up to a gauge transformation. 
The little-group representation $\mathcal{L}$ for massless particles realises non-trivially only the $SO(2)$ rotations $R(\theta)$ inside the Euclidean group $ISO(2)$ that leaves invariant the reference vector $k_r=(E,0,0,E)^T$ (in order to have finite dimensional representations). This is possible  iff $\sigma=B-A$ \cite{Weinberg:1995mt,Weinberg:1964ev}. For example, left- and right-handed Weyl representations $(1/2,0)$ and $(0,1/2)$ annihilate only particles of helicity $-1/2$ and $+1/2$ respectively, and create only anti-particles of helicity $1/2$ and $-1/2$ by $CPT$. Analogously, $(1,0)$ or $(0,1)$ (anti-symmetric self-dual tensors with two indexes) represent ordinary spin-1 fields, while the vector representation can not. In fact, any ordinary symmetric traceless representation $(A,A)$ for massless fields corresponds to spin-0 only.  
The density matrices of ordinary massless fields $(S,0)$ or $(0,S)$ can be written as monomials $P(p)$ of the 4-momentum $p$ of order $2S$  with $P(-p)=(-1)^{2S}P(p)$ \cite{Weinberg:1964ev}. Under $CPT$ the polarizations of particles and anti-particles map into each-other up to a sign as it happens in Eq.~(\ref{eq:generalCPTwave}) for the massive case.

In order to consider more general massless fields, such as the spin-1 photon field $A_\mu\sim (1/2,1/2)$, or the massless spin-2 graviton field $h_{\mu\nu}\sim (1,1)$, the notion of covariance must be extended by allowing gauge transformations, such as e.g. $U(\Lambda)A_\mu(x)U^\dagger(\Lambda)=(\Lambda^{-1})^\nu_\mu A_{\nu}(\Lambda x)+ \partial_\mu\omega(x,\Lambda)$. 
Indeed, the  polarizations in the helicity basis are defined  with respect to the little-group reference vector $k_r=(k,0,0,k)$ as the eigenvectors with largest eigenvalue, $\sigma=\pm S$, with respect to rotations around the $z$-axis. Thus, under the little-group, the analog of Eq.~(\ref{defuL}) for a massless spin-1 tells us that $\epsilon_\mu(k_r)$ shifts proportionally to the momentum, i.e.  $T^{\mu}_\alpha(\alpha,\beta) R^\alpha_\nu(\theta) \epsilon^{\nu\,\sigma}(\k_r)=e^{\pm i\sigma\theta}\left[\epsilon^{\mu\,\sigma}(\k_r)- (\alpha \pm i\beta)k_{r}^{\mu}\right]$,
 where the generic little-group transformation $W=T(\alpha,\beta)R(\theta)$ of $ISO(2)$ has been decomposed in a 2D translation $T(\alpha,\beta)$ and a 1D rotation $R(\theta)$.

\section{Linear polarizations and crossing symmetric amplitudes}\label{sec:crossforw}

A crossing symmetric amplitude can be written as the linear combination of amplitudes for particles/anti-particles of definite helicities, e.g. $\mathcal{M}(1^{\sigma_1}_{a_1} 2^{\sigma_2}_{a_2} \rightarrow 1^{\sigma_1}_{a_3} 2^{\sigma_2}_{a_2})+\mathcal{M}(1^{-\sigma_1}_{\overline{a}_3} 2^{\sigma_2}_{a_2} \rightarrow 1^{-\sigma_1}_{\overline{a}_1} 2^{\sigma_2}_{a_2})$.
For self-conjugate particles that are i.e. their own anti-particles, it may be useful to work instead with \textit{linear polarizations} that give rise to neat relations under crossing \cite{Bellazzini:2015cra,Cheung:2016yqr}. 
Consider for example a spin-1 particle described by a vector and define the linear basis of polarizations by the real 4-vectors $u_\ell^A=\epsilon^A_\mu(\k_r)=\delta^A_\mu$, that is
\begin{equation}
\label{eq:spin1linearwaves}
\epsilon_\mu^1(\k)=(0,1,0,0)^T\,,\qquad \epsilon_\mu^2(\k)=(0,0,1,0)^T \,, \qquad \epsilon_\mu^3(\k)=\frac{1}{m}(k^3,0,0,k^0)^T\,.
\end{equation}
The main advantage of this basis is that  the polarizations are real and the same for creation and annihilation, $v^A_\ell=\epsilon^{A\,*}_\mu=u_\ell$. Therefore, it is the amplitude of the crossed process with the \textit{same linear polarization} that is now related to the original scattering amplitude by exchanging $k\rightarrow -k$:
\begin{equation}
\mathcal{M}(\bar{1}^{A_1}_{\overline{a}_3} 2^{A_2}_{a_2}\rightarrow \bar{1}^{A_1}_{\overline{a}_1} 2^{A_2}_{a_4}, s)=
\mathcal{M}(1^{A_1}_{a_1} 2^{A_2}_{a_2}\rightarrow 1^{A_1}_{a_3} 2^{A_2}_{a_4}, u)\,.
\end{equation} 
In turn, a single amplitude for self-conjugate particles (carrying real representations of any internal symmetry group) with linear polarizations is itself, alone, already crossing symmetric in the forward elastic limit 
\begin{equation}
\mathcal{M}^{\mathrm{CS}}=\mathcal{M}(1^{A_1}_{a_1} 2^{A_2}_{a_2}\rightarrow 1^{A_1}_{a_1} 2^{A_2}_{a_2}, s)
\end{equation}
where we have chosen a real basis for the internal quantum number, $a_i=\overline{a}_{i}$.

The linear basis exists for any integer spin created/annihilated by a tensor representation with $n$ indexes: one can explicitly build the linearly polarizations by taking the the tensor product of $n$ copies of the $\epsilon^A_\mu$ in (\ref{eq:spin1linearwaves}),  and then decompose into the irreducible representations by symmetrization/anti-symmetrization and removing the traces. 
For spin-1/2 fermions, the Majorana basis of the $\gamma$-matrices is completely imaginary and one has $v=u^*$. From the Dirac equations, $(\slashed{p}-m)u(\p)=(\slashed{p}+m)v(\p)=0$,  the resulting polarizations can be taken real for $m=0$: the Majorana basis for massless particles is nothing but  the linear basis for the spin-1/2 polarizations that gives rise to crossing symmetric amplitudes.

 \section{Sum rules, lower subtractions and mixed states} 
 \label{sec:sumrules}
 
 In order to study $O(p^2)$ terms in the EFT one may be interested in the dispersion relations for $\mathcal{M}^\prime$:
 \begin{equation}
\mathcal{M}_{a_1a_2}^{\prime\,\sigma_1 \sigma_2}(\Lambda^2 \gg \mu^2 \gg m_i^2)\big|_{EFT} = \frac{1}{\pi}\left(\int^\infty_{s_{IR}} \frac{ds}{(s-\mu^2)^2} +\int_{-\infty}^{u_{IR}} \frac{ds}{(s-\mu^2)^2}\right) \mathrm{Im}\mathcal{M}^{\sigma_1\sigma_2}_{a_1 a_2}(s+i\epsilon)+C_{\infty}
\end{equation}
As the Froissart bound \cite{Froissart:1961ux} is no longer enough to discard $C_{\infty}$, one needs to make extra assumption about the UV behavior of the amplitude, see e.g. \cite{Bellazzini:2014waa}. 
  But even discarding $C_\infty$ (or claiming its finiteness and positivity), the extra assumptions are required also to derive positivity from the sum rules 
  \begin{align}
&\mathcal{M}_{a_1a_2}^{\prime\,\sigma_1 \sigma_2}(\Lambda^2 \gg \mu^2 \gg m_i^2)\big|_{EFT} =  \\
\nonumber
& \frac{1}{\pi}\int^\infty_{s_{IR}} \frac{ds}{(s-\mu^2)^2} \Pi(s) \sigma^{\mathrm{tot}}(1_{a}^{\sigma_1}  2_{a_2}^{\sigma_2})(s)  - \frac{1}{\pi}\int_{s_{IR}}^{\infty} \frac{ds}{(s-2(m_1^2+m_2^2)+\mu^2)^3} \Pi(s)\sigma^{\mathrm{tot}}(\bar{1}_{\bar{a}}^{-\sigma_1}  2_{a_2}^{\sigma_2})(s) 
\end{align}
implied by unitarity and crossing symmetry because the two contributions from $s$- and $u$-channel enter with opposite sign.  The function $\Pi(s)$, which asymptotically goes like $s$, is the square-root that shows up in the optical theorem (\ref{Eq:opticalth}). With these extra assumptions the sum rules allow to obtain interesting results e.g. in composite Higgs models \cite{Bellazzini:2014waa,Low:2009di,Falkowski:2012vh,Urbano:2013aoa} as well as on dimension-6 four-Fermi operators with no derivatives \cite{Adams:2008hp}.
 
Alternatively, one could take the point of view where the sum rules are checked against the calculable contribution on the right-hand side due to the observed (or possibly observable) resonances exchanged in the scattering. 
This is e.g. one way to use the sum rules for the pions in low-energy QCD.  In this logic, it may prove useful to work with density matrices for actual mixed states since the polarizations are not always known or measurable in practice. Multiplying the pure states density matrices with the distributions $p_{\sigma_1}$ and $p^\prime_{\sigma_2}$ for the polarizations one gets sum rules for the total cross-sections $\overline{\sigma}^{\mathrm{tot}}(1_{a}  2_{a_2})$ and $\overline{\sigma}^{\mathrm{tot}}(\bar{1}_{\bar{a}}  2_{a_2})$ which are averaged over the initial state polarizations: 
  \begin{align}
&\sum_{\sigma_1\,\sigma_2} p_{\sigma_1} p^\prime_{\sigma_2}\mathcal{M}_{a_1a_2}^{\prime\,\sigma_1 \sigma_2}(\Lambda^2 \gg \mu^2 \gg m_i^2)\big|_{EFT} =  \\
\nonumber
& \frac{1}{\pi}\int^\infty_{s_{IR}} \frac{ds}{(s-\mu^2)^2} \Pi(s) \overline{\sigma}^{\mathrm{tot}}(1_{a}  2_{a_2})(s)  - \frac{1}{\pi}\int_{s_{IR}}^{\infty} \frac{ds}{(s-2(m_1^2+m_2^2)+\mu^2)^3} \Pi(s)\overline{\sigma}^{\mathrm{tot}}(\bar{1}_{\bar{a}}  2_{a_2})(s) 
\end{align}



\end{document}